 \documentclass[11pt,oneside,english]{amsart}

\usepackage[dvips]{graphicx}            
\usepackage{pslatex}

\usepackage[T1]{fontenc}
\usepackage{epsfig,url}
\usepackage[latin9]{inputenc}
\usepackage{geometry}
\usepackage{amsthm}
\usepackage{enumitem}
\usepackage{amstext}
\usepackage{amssymb}
\usepackage{setspace}

\usepackage{pstricks}
\usepackage{epsf}
\usepackage{epsfig}
\usepackage{feynarts}

\makeatletter
\numberwithin{equation}{section}
\numberwithin{figure}{section}
\theoremstyle{plain}
\newtheorem{thm}{\protect\theoremname}
  \theoremstyle{definition}
  \newtheorem{example}[thm]{\protect\examplename}
  
  \newtheorem{rem}[thm]{Remark}

\font \rus= wncyr10
\newcommand{\Pro}{\mathbb{P}}
\newcommand{\ZZ}{\mathbb{Z}}
\newcommand{\QQ}{\mathbb{Q}}

\newcommand{\CC}{\mathbb{C}}
\newcommand{\RR}{\mathbb{R}}

\newcommand{\NN}{\mathbb{N}}
\newcommand{\Mod}{\mathcal{M}}
\newcommand{\Ao}{\mathcal{A}}
\newcommand{\AF}{\,{}^F \!\!\mathcal{A}}
\newcommand{\AFbar}{\,{}^F \!\!\bar{\mathcal{A}}}
\newcommand{\id}{\mathrm{id}}

\newcommand{\shuffle}{\, \hbox{\rus x} \,}

\makeatother

\usepackage{babel}
  \providecommand{\examplename}{Example}
\providecommand{\theoremname}{Theorem}

\begin{document}

\begin{flushright}
MaPhy-AvH/2014-10
\end{flushright}

\title{Feynman integrals and iterated integrals on moduli spaces of curves of genus zero}

\author{Christian Bogner and Francis Brown}
\begin{abstract} This paper describes algorithms for the exact symbolic computation of period integrals on moduli spaces $\Mod_{0,n}$
of curves of genus $0$ with $n$ ordered marked points, and  applications to the computation of Feynman integrals.
\end{abstract}
\maketitle

\section{Introduction}

Let $n\geq 0$ and let $\Mod_{0,n}$ denote the moduli space of Riemann spheres with $n$ ordered marked points.
The main examples of  periods of $\Mod_{0,n+3}$ consist of  integrals \cite{Bro2, Bro3, Ter}
\begin{equation}  \label{introM0nint}
\int_{0\leq t_1 \leq \ldots \leq t_n \leq 1} {\prod_{i=1}^n t_i^{a_{i}} (1-t_i)^{b_i}  \over \prod_{1 \leq i< j \leq n} (t_i-t_j)^{c_{ij}} } dt_1 \ldots dt_n   
\end{equation}
for suitable choices of integers $a_i,b_i,c_{ij} \in \ZZ $ such that the integral converges. These integrals have a variety of applications ranging 
from superstring theory \cite{Sch1, Sch2} to irrationality proofs \cite{AperyVar, Fisch1}. 
In \cite{Bro2} it was shown that such integrals  are linear combinations of multiple zeta values
\begin{equation} \label{introMZVdef}
\zeta(n_1,\ldots, n_r)  = \sum_{1\leq k_1 < \ldots < k_r} {1 \over k_1^{n_1} \ldots k_r^{n_r}} \qquad \hbox{ where }  n_i \in \NN, n_r \geq 2 
\end{equation}
 with rational coefficients. One of the goals of this paper  
is to provide effective algorithms, based on \cite{Bro2},  for computing such integrals $(\ref{introM0nint})$  symbolically.  The idea is to integrate out 
one variable at a time by working in a  suitable algebra of  iterated integrals (or rather, their symbols) which is closed under the two operations 
of taking primitives and taking limits along boundary divisors.

The second main application is  for 
the calculation of a large class of Feynman amplitudes, based on the universal property of the spaces $\Mod_{0,n}$. 
The general idea goes as follows. Suppose that $X \rightarrow S$ is a stable curve of genus zero. 
Then the universal property of moduli spaces yields an $n\geq3$ and  a commutative diagram:
\begin{equation} \label{Square}
\begin{array}{ccc} 
 X &  \longrightarrow  & \overline{\Mod}_{0,n+1}   \\
 \downarrow  &   & \downarrow   \\
 S &  \longrightarrow  &    \overline{\Mod}_{0,n} 
\end{array}
\end{equation}
The idea is that, for a specific class of (multi-valued) forms on $X$, we can integrate in the fibers of $X$ over $S$ by passing  to the right-hand side of the diagram
and computing the integral on the moduli space $\Mod_{0,n+1}$.   In this way, it  only suffices to  describe algorithms to integrate on the universal curve  $\Mod_{0,n+1}$
over $\Mod_{0,n}$. In practice, this involves computing a change of variables to pass from $X$ to a set of convenient coordinates on the  moduli space $\Mod_{0,n+1}$, applying the algorithm of \cite{Bro2} to integrate out one of these coordinates,
and finally  changing variables back to $S$.

This process can be repeated for certain varieties which can be fibered in curves of genus $0$ and yields an effective algorithm 
for computing a large class of integrals. Necessary conditions for such fibrations to exist  (`linear reducibility') were described in \cite{Bro4} and 
apply to many families of Feynman integrals, as we discuss in more detail presently.

\subsection{Feynman integrals}
Any Feynman integral in  even-dimensional space-time can  always be expressed as an integral in Schwinger parameters $\alpha_j$: 
\begin{equation}  \label{introFeynI}
I = \int_{0\leq \alpha_j \leq \infty } {P(\alpha_j) \over Q(\alpha_j)} \,  d\alpha_1 \ldots d{\alpha_N}   
\end{equation} 
 where $P$ and $Q$ are polynomials with (typically) rational coefficients and which perhaps depend on other parameters such as masses or momenta.
Cohomologial considerations tell us  that the types of numbers occurring as such  integrals  only depend on the denominator $Q$, and not on the numerator $P$. 
 A basic idea of \cite{Bro5} is to compute the integral $(\ref{introFeynI})$ by integrating out the Schwinger parameters $\alpha_i$ one at a time in some well-chosen order.
 After $i$ integrations, we require that the partial integral
  \begin{equation} \label{Ikpartial} 
  I(\alpha_1, \ldots, \alpha_{N-i}) = \int_{0\leq \alpha_j \leq \infty} {P \over Q}\, d\alpha_{N-i+1} \ldots d \alpha_{N}
  \end{equation}
   be expressed 
 as a certain kind of generalised polylogarithm function, or  iterated integral. Under certain conditions on the  singularities  of the integrand, the next variable can be integrated out.
  A `linear reduction' algorithm \cite{Bro4,Bro5}  yields   an upper bound for the set of singularities of   
 $(\ref{Ikpartial})$ and can tell us in advance whether  $(\ref{introFeynI})$ can be computed by this method.  It takes the form of a sequence of sets of polynomials (or rather,  their associated hypersurfaces):
 $$S_1 \ , \ S_2  \ , \  \ldots $$
 where $S_1 = \{Q\}$, and $S_{i+1}$ is derived from $S_{i}$ by taking certain resultants of polynomials in $S_i$ with respect to $\alpha_{N-i+1}$. 
When $Q$ is linearly reducible, we obtain a sequence of spaces  for $i\geq 1$:
\begin{eqnarray} 
X_i & =&   (\Pro^1 \backslash \{0,\infty\})^{N-i+1} \ \backslash \ V(S_i)  \nonumber \\ 
 & =& \{ (\alpha_{1}, \ldots, \alpha_{N-i+1}):  \alpha_k \neq 0, \infty \hbox{ and } P(\alpha_{1},\ldots, \alpha_{N-i+1}) \neq 0 \hbox{ for all } P \in S_i \} \nonumber 
\end{eqnarray}
and maps $\pi_i: X_i \rightarrow X_{i+1}$ which correspond to projecting out the variable $\alpha_{N-i+1}$. 
The linear reducibility assumption guarantees that $X_i$ fibers over $X_{i+1}$ in  curves of genus $0$. Thus  setting $(X,S) = (X_i, X_{i+1})$ in the discussion above, we can explicitly find changes of variables in the $\alpha_i$ to write $(\ref{Ikpartial})$ as an iterated integral on a moduli space $\Mod_{0,n}$ and do the next integration.

It is perhaps surprising that such a method should ever work for any non-trivial Feynman integrals. The fundamental reason it does, however,
is that the polynomial $Q$ can be expressed in terms of determinants of matrices whose entries are linear in the $\alpha_i$ parameters.
In the case when $Q$ is the first Symanzik polynomial, and to a  lesser extent when $Q$ also depends on masses and external momenta,  it satisfies
many `resultant identities', which only  break down at  a certain  loop order.

\subsection{A method of hyperlogarithms versus a  method of moduli spaces} \label{secthypversusmod}
There are two possible approaches to implementing the above algorithm: one which is now referred to as the `method of hyperlogarithms' \cite{Bro5},
which stays firmly on the left-hand side of the diagram $(\ref{Square})$;  the other, which is the algorithm described here  \cite{Bro2},  which makes more systematic use of 
the geometry of the moduli spaces $\Mod_{0,n}$ and works on the right-hand side of the diagram $(\ref{Square})$.

The first involves working directly in Schwinger parameters, and expressing all partial integrals as hyperlogarithms  (iterated integrals of one variable) 
whose arguments are certain rational functions in Schwinger parameters. It has been fully implemented by Panzer \cite{Pan1,Pan2,Pan3} and various
parts of the algorithm have found applications in different contexts, as described below. A conceptual disadvantage of this method is that the underlying geometry of every Feynman diagram is different.

The second method, espoused here, is to compute all integrals on the moduli spaces $\Mod_{0,n}$ (which, by no accident, are the universal domain of definition for hyperlogarithms).  Thus the underlying geometry is always the same and is well-understood;  all the information about the particular integral $(\ref{introFeynI})$ is contained in the changes of variables $(\ref{Square})$.  Another key difference is the systematic use of generalised symbols of functions in several complex variables, as opposed to  functions of a single 
variable (hyperlogarithms).

 That these two points of view are equivalent is theorem \ref{thmVdecomp} below, but leads, in practice, to a rather different algorithmic approach. We nonetheless provide
 algorithms (the symbol and unshuffle maps) to pass between both points of view.

\subsection{Applicability}
The above method can be applied to a range of Feynman integrals provided that the initial integral $(\ref{introFeynI})$ is convergent.
The case of massless, single-scale, primitively overall-divergent Feynman diagrams in a scalar field theory was detailed in \cite{Bro4}. Since then, 
the method was applied to the computation of integrals of hexagonal  Feynman graphs, arising in $\mathcal{N}=4$ supersymmetric Yang-Mills theory \cite{Del1, Del2, Del3}, 
integrals with operator insertions contributing to massive matrix elements of quantum chromodynamics (QCD) \cite{Abl1, Abl2, Abl3}, 
one- and two-loop triangular Feynman graphs with off-shell legs \cite{Cha}, 
phase-space contributions \cite{Ana1, Ana2} to the cross-section for threshold production of the Higgs boson 
from gluon-fusion at N3LO QCD \cite{Higgs}, 
coefficients in the expansion of certain hypergeometric functions, contributing to superstring amplitudes \cite{Sch1, Sch2}, 
massless multi-loop propagator-type integrals \cite{Pan1}, 
and a variety of three- and four-point Feynman integrals depending on several kinematical scales \cite{Pan2}. 
These applications arise from very different contexts and the method is combined with various other computational techniques. Focussing on Feynman integrals, we can summarize by stating   that the method can be extended 
to the following situations:
\begin{itemize}
\item To Feynman graphs with several masses or kinematic scales.
\item To gauge theories, or more generally, integrals with arbitrary numerator structures.
\item To graphs with ultra-violet  subdivergences. In particular, it is compatible with the renormalisation procedure due to Bogoliubov, Parasiuk, Hepp and Zimmermann (BPHZ) in a momentum scheme \cite{BroKre}.
\item Finally, it can also  be combined with dimensional regularisation to treat UV and IR divergences by the method of \cite{Pan2}.
\end{itemize}
The method is suited for automatization on a computer. For the special case of harmonic polylogarithms, the programs \cite {Mai1, Mai2} support  direct integration using 
these functions. For the general approach, using hyperlogarithms, a first implementation of the method was presented in ref. \cite{Pan3}. A program for the numerical evaluation of these functions is 
given in ref. \cite{Vol}.

There appear to be other classes of integrals which are not strictly Feynman diagrams, but for which the method of iterated fibration in curves of genus  zero $(\ref{Square})$
still applies. A basic example are periods of arbitrary hyperplane complements \cite{Bro2}, and as a consequence, various families of integrals occurring in deformation quantization, for example.

\subsection{Plan of the paper} In section $\ref{Sect2}$ we review some of the mathematics of iterated integrals on moduli spaces  $\Mod_{0,n}$, based on
\cite{Bro2}. The geometric ideas behind the main algorithms are outlined here. In \S\ref{sec:Computing-on-the}, these algorithms are spelled out in complete detail
together with some illustrative examples. 
In  \S\ref{sec:Feynman-type-integrals}, it is explained how to pass between Feynman integral representations and moduli space representations.
In \S\ref{sec:Applications} we discuss some applications, before presenting the conclusions. Some introductory background can be found in the survey papers \cite{Bro6,Bro7}. 
\\

The methods of  $\S\ref{sec:Computing-on-the}$  should in principle generalise to genus $1$, using multiple elliptic polylogarithms defined in \cite{BrLe}, but there remains
a considerable amount of theoretical groundwork to be done.
A different direction for generalisation is to introduce roots of unity, by replacing $\Pro^1\backslash \{0,1,\infty\}$ with $\Pro^1 \backslash \{0,\mu_N, \infty\}$ where $\mu_N$
is the group of $N^\mathrm{th}$ roots of unity. This should be rather similar to the framework discussed here.
\\

\emph{Acknowledgements}: The second named author is a beneficiary of ERC grant 257638. The first named author thanks Erik Panzer for very useful discussions and especially helpful suggestions regarding the contents of section \ref{sec:Feynman-type-integrals}.
We thank Humboldt University for hospitality and support. Our Feynman graphs were drawn using \cite{Hah}.

\section{Iterated integrals on the moduli spaces $\Mod_{0,n}$} \label{Sect2}

\subsection{Coordinates}
Let $n\geq 3$ and let $\CC_{\infty}= \CC \cup \{\infty\}$ denote the Riemann sphere.  The  complex moduli space $\Mod_{0,n}(\CC)$ is  the space of 
$n$ distinct ordered points on $\CC_{\infty}$ modulo automorphisms 
$$\Mod_{0,n}(\CC) = \{ (z_1,\ldots, z_n)\in \CC_{\infty}^n \hbox{ distinct} \} /\mathrm{PGL_2}(\CC) \ . $$
There are two sets of coordinates,  called simplicial and cubical, which are useful for the sequel. By applying an element of $\mathrm{PGL_2}(\CC)$, we can 
assume that $z_1 =0, z_{n-1} =1$ and $z_n = \infty$ and define 
$$t_1 = z_2 \ , \ t_2 = z_3 \ ,  \ldots\ , \  t_{n-3} = z_{n-2} \ .$$
The $(t_1,\ldots, t_{n-3})$ are called  simplicial coordinates and define an isomorphism
$$\Mod_{0,n}(\CC) \cong \{(t_1,\ldots, t_{n-3}) \in \CC^{n-3} \hbox { such that the } t_i \hbox{ are distinct and }  t_i \neq 0,1\}\ .$$
Cubical coordinates, on the other hand, are defined by 
\begin{equation}  \label{simplicialtocube}
  x_1 = {t_1 \over t_2} \ , \ x_2 = {t_2 \over t_3 } \ ,  \ \ldots\ , \  x_{n-4} = {t_{n-4} \over t_{n-3}} \ , \ x_{n-3}= t_{n-3}  
  \end{equation}
Cubical coordinates define  an isomorphism
$$\Mod_{0,n}(\CC) \cong \{(x_1,\ldots, x_{n-3}) \in \CC^{n-3} \hbox { such that } x_{i} x_{i+1} \ldots x_j  \neq \{0,1\}  \hbox{ for all  }  1 \leq i \leq j \leq n-3\}\ .$$
Note that the divisors above  only involve products of cubical coordinates with consecutive indices.  The main advantage of cubical coordinates is that 
the divisors corresponding to  
 $$x_i=0  \quad \hbox{for } i=1,\ldots, n-3 $$
are strict normal crossing 
in a neighbourhood of the  origin $(0,\ldots, 0)$. The reason for the nomenclature is that the standard cell (a connected component of the set of real points
$\Mod_{0,n}(\RR)$) is either a simplex:
$$ X_n \cong \{(t_1,\ldots, t_{n-3}) \in \RR^{n-3}: 0 < t_1 < \ldots < t_{n-3} < 1\}$$
or a cube:
$$
 X_n \cong \{(x_1,\ldots, x_{n-3}) \in \RR^{n-3}: 0 <  x_i < 1 \hbox{ for all } 1\leq i \leq n-3 \}\  , $$
depending on the choice of coordinate system.

\subsection{Differential forms} Let $\Omega^k(\Mod_{0,n})$ denote the space of  
global regular differential $k$-forms on $\Mod_{0,n}$ which are defined over $\QQ$. 
Consider  the following elements of  $\Omega^1(\Mod_{0,n})$:
$$\omega_{ij} = { {dt_i - dt_j}  \over t_i -t_j } \ \hbox{ for }  \ 0\leq i,j\leq n-2$$
where we set $t_0=0$ and $t_{n-2}= 1$. Clearly $\omega_{ij} = \omega_{ji}$ and $\omega_{ii}=0$. There are no other linear relations between the $\omega_{ij}$ besides these.
Define 
$$\Ao^1(\Mod_{0,n}) = \langle \omega_{ij}:  \hbox{ for  }  i < j  \ ,  \  (i,j) \neq (0,n-2)\rangle_{\QQ} $$
Thus  $\Ao^1(\Mod_{0,4})$ has the basis ${dt_1 \over t_1}, {dt_1 \over t_1 -1}$.
The $\omega_{ij}$ satisfy the following quadratic relation:
\begin{equation} \label{quadrel}
\omega_{ij} \wedge \omega_{jk}+ \omega_{jk} \wedge \omega_{ki}
+ \omega_{ki} \wedge \omega_{ij}
=0
\end{equation}
for all indices $i,j,k$. Define $\Ao^{\bullet}(\Mod_{0,n})$ to be the differential graded algebra which is the quotient of the  exterior algebra
generated by $\Ao^1(\Mod_{0,n})$ by the quadratic relations $(\ref{quadrel})$. A theorem due to Arnold states that  
$$\Ao^{\bullet}(\Mod_{0,n}) \longrightarrow H_{dR}^{\bullet}(\Mod_{0,n};\QQ)$$
is an isomorphism of algebras. Thus $\Ao^{\bullet}(\Mod_{0,n})$ is an explicit  model for the de Rham cohomology of $\Mod_{0,n}$.
In cubical coordinates, it is convenient to  take a different basis for $\Ao^1(\Mod_{0,n})$ formed by 
$${dx_i \over x_i} \quad  \hbox{ and } \quad   { { d (x_i\ldots x_j) \over x_i x_{i+1} \ldots x_j -1 }    } \ \hbox{ for }  1\leq i \leq j \leq n-3 \ .  $$
We will consider iterated integrals in these one-forms.

\subsection{Iterated integrals and symbols}
Recall the definition of iterated integrals from \cite{Che}. Let $M$ be a smooth complex manifold and let $\omega_1, \ldots, \omega_n$
denote smooth 1-forms. Let $\gamma:[0,1] \rightarrow M$ be a smooth path. The iterated integral of these forms  along $\gamma$ is  defined by
$$\int_{\gamma} \omega_1 \ldots \omega_n = \int_{0\leq t_1 \leq t_2 \leq \ldots \leq t_n \leq 1} \gamma^{*}(\omega_n) (t_1) \ldots \gamma^{*}(\omega_1) (t_n)\ .$$
There are different conventions for iterated integrals: here we integrate starting from the right.
The argument of the left-hand integral is $\CC$-multilinear  in the forms  $\omega_i$ and can be viewed as a functional
on the tensor product $\Omega^1(M)^{\otimes n}$. 
Elements of this space are customarily written using the 
bar notation
$[\omega_1 | \ldots | \omega_n]$ to denote a tensor product $\omega_1 \otimes \ldots \otimes \omega_n$.

Chen's theorem states that iterated integration defines an isomorphism from the zeroth cohomology  of the reduced bar construction 
on the $C^{\infty}$ de Rham complex  of $M$ to the space of iterated integrals on $M$ which only depend on the homotopy class of $\gamma$ relative to its endpoints.
The reduced bar construction on $\Mod_{0,n}$ can be written down explicitly using the model $\Ao$ defined above, in terms of a certain algebra of symbols.
For $n\geq 3$,  define a graded $\QQ$ vector space
$$V(\Mod_{0,n}) \subset    \bigoplus_{m\geq 0} \Ao^1(\Mod_{0,n})^{\otimes m} $$
by linear combinations of bar elements
$$   \sum_{I=(i_1,\ldots, i_m)}  c_I [\omega_{i_1} | \ldots | \omega_{i_m}]  $$
which satisfy the integrability condition
\begin{equation}\label{intcond} 
\sum_I c_I [\omega_{i_1} | \ldots  |   \omega_{i_{j-1}} | \omega_{i_j} \wedge \omega_{i_{j+1}} |\omega_{i_{j+2}}| \ldots | \omega_{i_m}] = 0  \qquad  \hbox{ for all } 1\leq j \leq m-1\ . 
\end{equation} 
Then $V(\Mod_{0,n})$  is an algebra for the 
shuffle product $\shuffle$ and is equipped with the deconcatenation coproduct $\Delta$, which is defined by: 
$$\Delta [\omega_{i_1} | \ldots | \omega_{i_m}]   = \sum_{k=0}^m  [\omega_{i_1} | \ldots | \omega_{i_k}] \otimes  [\omega_{i_{k+1}} | \ldots | \omega_{i_m}]$$
 Thus    $V(\Mod_{0,n})$ is a graded Hopf algebra over $\QQ$. Iterated integration defines a homomorphism
\begin{eqnarray}
V(\Mod_{0,n})  &\longrightarrow&   \{\hbox{Multivalued functions  on } \Mod_{0,n}(\CC) \}  \label{itintonMod0n} \\
\sum_{I=(i_1,\ldots, i_m)} c_I [ \omega_{i_1} | \ldots | \omega_{i_m}]   & \mapsto & \sum_I c_I \int_{\gamma_z}   \omega_{i_1}  \ldots  \omega_{i_m}   \nonumber 
\end{eqnarray}
where $\gamma_z$ is a homotopy equivalence class of paths from a fixed (tangential) base-point  to $z \in \Mod_{0,n}(\CC)$. 
By a version of Chen's theorem, this map gives an isomorphism   between  homotopy invariant iterated integrals (viewed as  multi-valued functions of their endpoint) on $\Mod_{0,n}$ and symbols. 
Equivalently, this means that the map $(\ref{itintonMod0n})$ is a homomorphism of differential algebras (for a certain differential to be defined in $(\ref{BdRdifferential})$)
and the constants of integration are fixed as follows.
One can show that, in cubical coordinates $(x_1,\ldots, x_{n-3})$, every iterated integral $(\ref{itintonMod0n})$  admits  a finite  expansion
of the form
$$\sum_{I=(i_1,\ldots, i_{n-3})}    f_I(x_1,\ldots, x_{n-3}) \log(x_1)^{i_1} \ldots \log(x_{n-3})^{i_{n-3}} $$
where $f_I(x_1,\ldots, x_{n-3})$ is a formal power series in the $x_i$ which converges in the neighbourhood of the origin. The normalisation condition is that
the regularised value at zero vanishes:
$$f_{0,\ldots, 0} (0,\ldots, 0) = 0\ .$$
This gives a bijection between symbols and certain multivalued  functions (whose branch is fixed, for example, on the standard cell $X_n)$, and in this way we can work entirely with symbols. Various operations on functions  can be 
expressed algebraically in terms of $V(\Mod_{0,n})$. For example, the monodromy of functions around loops can be expressed in terms of the coproduct $\Delta$.

\subsection{The bar-de Rham complex} \label{sectbardeRham}
Differentiation of iterated integrals with respect to their endpoint corresponds to the following left-truncation operator
\begin{eqnarray} \label{BdRdifferential}
d: V(\Mod_{0,n})  & \longrightarrow & \Omega^1(\Mod_{0,n}) \otimes   V(\Mod_{0,n})\\
\sum_I c_I [ \omega_{i_1} | \ldots | \omega_{i_m}]  & \mapsto &  \sum_I  c_I \omega_{i_1} \otimes [ \omega_{i_2}  | \ldots | \omega_{i_m}]  \nonumber 
\end{eqnarray} 
where $I= (i_1,\ldots, i_m)$. 
The bar-de Rham complex  is defined to be
$$ B(\Mod_{0,n}) = \Omega^{\bullet}(\Mod_{0,n}) \otimes V(\Mod_{0,n})$$
equipped with the differential induced by $d$. In \cite{Bro2} it was shown that 
\begin{thm} The cohomology of the bar-de Rham complex of $\Mod_{0,n}$ is trivial:
$$H^i (  B(\Mod_{0,n})) = \begin{cases} \QQ \quad  \hbox{ if } i = 0 \nonumber \\ 0 \quad \hbox{ if } i > 0 \end{cases}$$
\end{thm}
In particular,  $B(\Mod_{0,n})$ is closed under the operation of taking primitives, which is one ingredient for computing integrals symbolically.
The next ingredient states that  one can compute regularised limits along  irreducible boundary divisors $D \subset \overline{\Mod}_{0,n} \backslash \Mod_{0,n}$
with respect to certain local canonical  sections $v$ of the normal bundle of  $D$. Let $\mathcal{Z}$ denote the $\QQ$-vector space generated by multiple zeta values 
$(\ref{introMZVdef}).$
\begin{thm}  \label{thmreglimmap} There exist   canonical `regularised limit' maps 
$$\mathrm{Reg}^v_D : V(\Mod_{0,n}) \longrightarrow V(\Mod_{0,r}) \otimes V(\Mod_{0,n+2-r}) \otimes \mathcal{Z}$$
for every irreducible boundary divisor $D$ of $\overline{\Mod}_{0,n}$ which is isomorphic to $\overline{\Mod}_{0,r} \times \overline{\Mod}_{0,n+2-r}$.
\end{thm}
This states that the regularised limits of iterated integrals on moduli spaces are products of such iterated integrals with coefficients in the ring $\mathcal{Z}$ of multiple zeta values.
By applying these two operations of primitives and limits, one can compute period integrals on $\Mod_{0,n}$.   In more detail:

\subsubsection{Total primitives}  \label{sectTotalPrimitives} Taking primitives of differential one-forms is a trivial matter. 
Let $\eta $ be a $1$-form in  $B^1(\Mod_{0,n})$ such that $d\eta=0$.  We can write it as a finite sum
$$\eta = \sum_k \omega^{k}_{0} \otimes  [ \omega_1^k| \ldots | \omega_n^k] $$
A primitive is given explicitly by
$$\int \eta = \sum_k [ \omega^{k}_{0}| \omega_1^k| \ldots | \omega_n^k] \ .$$
The constant of integration is uniquely (and automatically) determined by the property 
$$\varepsilon(\int \eta ) =0 $$
where $\varepsilon: V(\Mod_{0,n}) \rightarrow \QQ$ is the augmentation map (projection onto terms of weight $0$). 
The fact that $\int \eta $  satisfies the integrability condition $(\ref{intcond})$ follows from the integrability of $\eta$ and the equation $d\eta=0$.
In practice, the algorithm we actually use for taking primitives on the universal curve needs to be  more  sophisticated and is described below.

\subsubsection{Limits} 
When taking limits, one must bear in mind the fact that 
the elements of $V(\Mod_{0,n})$ represent multivalued functions, and 
hence depend on the (homotopy class) of the path $\gamma_z$ of analytic continuation $(\ref{itintonMod0n})$.  When computing period integrals by the method
described above,  however,  all iterated integrals which occur will be single-valued on the domain of integration (\cite{Bro4}, theorem 58). 

In cubical coordinates, the domain of integration is the unit cube $X_n = [0,1]^{n-3}$, and so it suffices in this case to define limits
along the divisors in $\overline{\Mod}_{0,n}$ defined by $x_i=0$ and $x_i=1$, where $x_i$ are cubical coordinates.
Recall that the integration map from $V(\Mod_{0,n})$ to multivalued functions is normalised at the point $(0,\ldots, 0)$ with respect to unit tangent vectors in  cubical 
coordinates $x_i$, and it follows that the limits at $x_i=0$ are trivial to compute.  Any  function $f$ in the image of $(\ref{itintonMod0n})$ is uniquely determined on 
the simply connected domain $X_n=[0,1]^{n-3}$, and  admits a unique expansion for some $N$
\begin{equation} \label{fexpansion} 
f(x_1,\ldots, 1-\epsilon_i,\ldots, x_{n-3}) = \sum_{k=0}^N  \log(\epsilon)^k p_k(\epsilon)  f_k(x_1,\ldots, x_{i-1},  x_{i+1}, \ldots,x_{n-3}) 
\end{equation} 
where $p_k(\epsilon)$ is holomorphic  at $\epsilon=0$ and where $f_k$ is in the image of $V(\Mod_{0,i+2}) \otimes V(\Mod_{0,n-i})$. 
The `regularised limit' of $f$ along $x_i=1$ (with respect to the normal vector $-{\partial \over \partial x_i}$) is the function
$$   \mathrm{Reg}_{x_i=1}\, f =  p_0(0)  f_0(x_1,\ldots, x_{i-1},  x_{i+1}, \ldots,x_{n-3})  \ .$$
It is the composition  of the realisation map $(\ref{itintonMod0n})$ with  a certain map (theorem $\ref{thmreglimmap}$)
$$V(\Mod_{0,n}) \longrightarrow  \mathcal{Z} \otimes V(\Mod_{0,i+2}) \otimes  V(\Mod_{0,n-i})$$
where $\mathcal{Z}$ is the ring of multiple zeta values.  This map can be computed explicitly as follows.

Recall first of all the general formula for the behaviour of iterated integrals with respect to composition of paths, where $\gamma_1 \gamma_2$ denotes the path
$\gamma_2$ followed by the path $\gamma_1$:
\begin{equation}  
\int_{\gamma_1 \gamma_2} \omega_1 \ldots \omega_n  = \sum_{i=0}^n \int_{\gamma_1} \omega_1 \ldots \omega_i \int_{\gamma_2} \omega_{i+1} \ldots \omega_n\  .\label{pathconcat}
\end{equation} 
 If $E_{\gamma}$ is the function on $V(\Mod_{0,n})$ which 
denotes evaluation of a (regularised) iterated integral along a path $\gamma$,  then  the previous equation can be interpreted as a convolution product:
\begin{equation} \label{convolution} 
E_{\gamma_1 \gamma_2} = m ( E_{\gamma_1} \otimes E_{\gamma_2}) \circ \Delta 
\end{equation} 
Ignoring, for the time being, issues to do with tangential base points and  regularisation, a path from the origin $0$ to a point $z=(x_1,\ldots,x_{i-1}, 1, x_{i+1}, \ldots, x_{n-3})$  which lies inside  the cube $X_n=[0,1]^{n-3}$ 
is homotopic to a composition of paths $\gamma_1 \gamma_2$
 (`up the $i^{\mathrm{th}}$ axis and then along to the point $z$'), where
$$\gamma_2  =\hbox{straight line from } \ 0\  \hbox{ to } \   1_i =( \underbrace{0,\ldots, 0}_{i-1}, 1, \underbrace{0,\ldots 0}_{n-i-4})$$
and $\gamma_1$ is a path from $1_i$ to $z$ which lies inside $x_i=1$. The segment of path $\gamma_2$ can be interpreted as a straight line
from $0$ to $1$ in $\Mod_{0,4} = \Pro^1\backslash \{0,1,\infty\}$ (with coordinate $x_i$). Iterated integrals along this  path  give rise to coefficients of the Drinfeld associator, which are  multiple zeta values.
Iterated integrals along $\gamma_1$ can be identified with our class of multivalued functions on the boundary divisor $D$
of $\overline{\Mod}_{0,n}$ defined by $x_i=1$, which is canonically isomorphic to $\overline{\Mod}_{0,i+2} \times \overline{\Mod}_{0,n-i}$. 
One can check that the  above argument makes sense for regularised (divergent) iterated integrals, and 
putting the pieces together yields the regularisation algorithm which is described below.

\begin{rem} For the computation of period integrals, one needs slightly more. We actually require 
an expansion  of the function $(\ref{fexpansion})$ as  a polynomial in $\log(\epsilon)$ and a Taylor expansion of $p_k(\epsilon)$  up to some order $K$ in  $\epsilon$.
This is because $f$ can occur with a rational prefactor which may have poles in $\epsilon$ of order $K$.  This Taylor expansion is straightforward to compute recursively  by expanding 
 ${\partial \over \partial x_i} f$ and integrating (we know the constant terms by the previous discussion).  The partial derivative   ${\partial \over \partial x_i} f$  is  simply a component of the total differential $d$ defined in $(\ref{BdRdifferential})$, which decreases the length and hence this gives an algorithm which  terminates after finitely many steps, 
 also described below.
\end{rem}

Note that in order to compute period integrals $(\ref{introM0nint})$, one only requires taking limits with respect to the  final cubical variable $x_i$ for $i=n$.

\subsubsection{More general limits}
It  can happen, for example when computing Feynman integrals, that one wants to take limits at more general divisors on $\overline{\Mod}_{0,n}$. 
The compactification of the standard cell $X_n$ (the closure of  $X_n$  in $\overline{\Mod}_{0,n}$ for the analytic topology) is a closed polytope 
$$\overline{X}_n \subset \overline{\Mod}_{0,n}$$
which has the combinatorial structure of a Stasheff polytope. 
It can happen  that one wants to compute limits at a (tangential) base point on the boundary of $\overline{X}_n$. An example is illustrated in the figure
below in the case $n=5$, and where $\overline{X}_5$ is a pentagon. 

\begin{figure}[h!]
  \begin{center}
   \epsfxsize=12cm \epsfbox{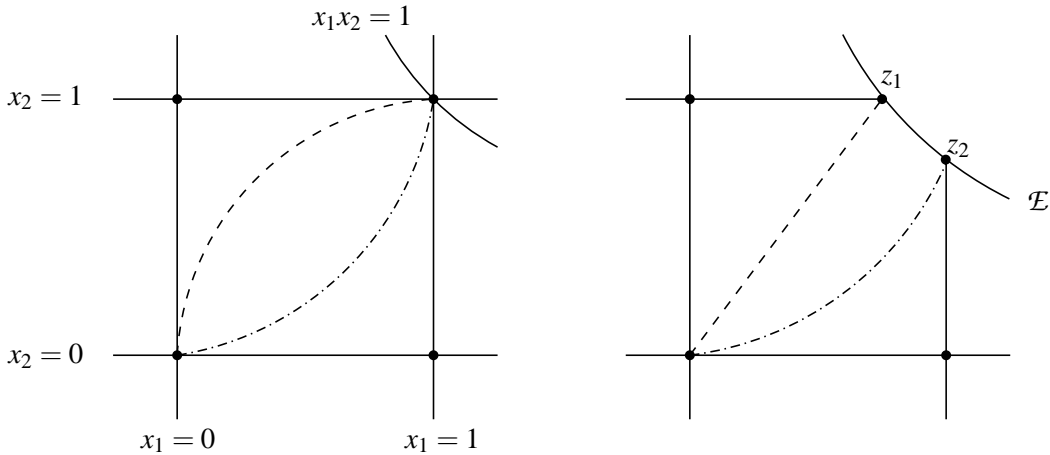}
         \put(-330,-10){$x_1=0$}  \put(-230,-10){$x_1=1$}
           \put(-380,22){$x_2=0$}  \put(-380,120){$x_2=1$}
            \put(-265,150){$x_1x_2=1$}  
     \put(5,80){$\mathcal{E}$} \put(-50,128){$z_1$}
     \put(-25,102){$z_2$}
 \caption{On the left is a picture of $\Mod_{0,5}$ in cubical coordinates $(x_1,x_2)$, and  two  paths going from the origin to $(1,1)$. On the right-hand side is the space
 obtained by blowing up the point $(1,1)$. The exceptional divisor is $\mathcal{E}\cong \Pro^1$. There are two tangential base points defined over $\ZZ$ which lie above $(1,1)$, which are based at $z_1$ and $z_2$. The inverse image  of  the  two paths end at the point $z_1$, or $z_2$ respectively.  
  }
  \end{center}
\end{figure}

The case of such limits can be dealt with using explicit local normal crossing coordinates on the boundary of $\overline{X}_n$ such as
the dihedral coordinates $u_{ij}$ defined in \cite{Bro2}. One can show that any such limit is in fact 
a composition of regularised limits along divisors $x_{i_k} =1$ and $x_{i_k}=0$ in some specified (but not necessarily unique) order. This order can be determined
from the combinatorics of the dihedral coordinates, and gives an algorithm to compute limits in this more general sense.

For example, in the figure, the point $z_1$ is reached by taking the limit first as $x_2$ goes to $1$ and then $x_1$ goes to $1$; the point $z_2$ corresponds
to the opposite order.      The regularised limits of iterated integrals (such as $\mathrm{Li}_{1,1}(x,y)$) at $(1,1)$  along each path are different.
 Note that a path which approaches $(1,1)$ with a gradient which is strictly in between $0$ and $\infty$ corresponds to a limit point 
which is not equal to either  $z_1$ or $z_2$ on  $\mathcal{E}$ and could take us outside the realm of multiple zeta values.

Finally, it is worth noting that one can imagine situations when one needs to take limits at points  `at infinity' corresponding to the case when,
for example, some cubical coordinates $x_i$ go to infinity. This will not be discussed here.

\subsection{Fibrations} The space $V(\Mod_{0,n})$ is defined by a  system of  quadratic equations $(\ref{intcond})$ and its structure is hard to
understand from this point of view. We will never need to actually solve the integrability equations $(\ref{intcond})$. 

A different description of $V(\Mod_{0,n})$ comes from considering the morphism
\begin{eqnarray}
\Mod_{0,n} & \longrightarrow & \Mod_{0,n-1} \label{fibration}  \\ 
(x_1,\ldots, x_{n-3}) & \mapsto & (x_1,\ldots, x_{n-4})  \nonumber
\end{eqnarray}
which is obtained by forgetting the last cubical coordinate.  It is a fibration, whose fiber over the point $(x_1,\ldots, x_{n-4})$ is isomorphic to 
the punctured projective line
$$C_{n} = \Pro^1 \backslash \{   0, (x_1\ldots x_{n-4})^{-1} , \ldots, x_{n-4}^{-1}, 1, \infty\} $$
with coordinate $x_{n-3}$.   Let $\Ao_n=\Ao(\Mod_{0,n})$ denote the model for the de Rham complex on $\Mod_{0,n}$ defined earlier, and let 
$\AFbar_n=  \Ao_{n}/\Ao_{n-1}$ denote the  $\QQ$-vector space of  relative differentials. 

Denote the natural projection  by
\begin{equation} \label{projectontoAf} \omega \mapsto \overline{\omega}: \Ao_{n} \rightarrow \AFbar_n
\end{equation}
Using the representation of forms in cubical coordinates, we can choose a splitting 
\begin{equation}  \label{lambdadefn}
\lambda_n \quad : \quad \AFbar_n \overset{\sim}{\rightarrow} \AF_n \subseteq \Ao_{n}
\end{equation} which is defined explicitly in $(\ref{lambdaexplicit})$, and obtain a decomposition of $\Ao_{n-1}$-modules:
\begin{equation} \label{Aosproductdecomp} 
\Ao_{n} \cong \Ao_{n-1} \otimes \AFbar_n\ . 
\end{equation}
Armed with this decomposition, the quadratic relation $(\ref{quadrel})$
can be reinterpreted as a multiplication law on $1$-forms on the fiber:
\begin{eqnarray} \label{Wedgedecomp} 
  \mu_n \quad : \quad  \AF_n^1 \wedge \AF_n^1 \longrightarrow \Ao^1_{n-1} \otimes \AF^1_n 
\end{eqnarray}
which is used intensively in all computations.    The product of two elements in $\AF^1_n$ lies in 
$\Ao_n^2 \cong \Ao^2_{n-1} \oplus (\Ao^1_{n-1} \otimes \AFbar^1_n)$ since $\AFbar_n^2=0$. In fact, our choice of splitting $\lambda_n$ is such that 
the component of the previous isomorphism in $\Ao_{n-1}^2$ vanishes, which defines the map $(\ref{Wedgedecomp})$.

\begin{thm} \cite{Bro2} \label{thmVdecomp} The choice of map $\lambda_n$  gives a canonical isomorphism of algebras 
\begin{equation} \label{Vdecomp}
  V(\Mod_{0,n}) \cong V(\Mod_{0,{n-1}}) \otimes V(C_n)\ , 
 \end{equation}
(which does not respect the coproducts on both sides) where 
$$V(C_n)  = \bigoplus_{k\geq 0} (\AFbar^1_n)^{\otimes k} $$
is  the $\QQ$-vector space spanned by all words in  $\AFbar^1_n$, equipped with the shuffle product. \end{thm}
This gives a very precise  description of the algebraic structure on  $V(\Mod_{0,n})$: by applying this theorem iteratively, every element of $V(\Mod_{0,n})$ can be uniquely represented by a  sum of tensors
of words in  prescribed alphabets.  In order to go back and forth between the two representations on the left and right hand sides of $(\ref{Vdecomp})$ 
we have the symbol and unshuffle maps, defined as follows.
\begin{enumerate}[itemsep=4pt,parsep=4pt]
\item The \emph{symbol map} is a homomorphism, which depends on the choice $(\ref{lambdadefn})$, 
\begin{equation}
\Psi: V(C_n)  \longrightarrow V(\Mod_{0,n})
\end{equation} 
which can be thought of as the map which takes a function defined on a fiber of the universal curve $C_n$ and extends it to a function on the entire moduli space $\Mod_{0,n}$.

It is constructed as follows. One can define a Gauss-Manin connection, corresponding to `differentiation under an iterated integral' which is a linear map
$$\nabla : V(C_n) \longrightarrow \Ao^1_{n-1} \otimes V(C_n)$$
by the following recipe: lift words in $\AFbar_n$ to words in $\Ao_{n}$ via the map $\lambda_n$; then apply the usual internal
differential   of the bar construction in degree $0$ (all signs simplify since the $\omega_i$ are $1$-forms):
\begin{equation} \label{integrability}
 D [\omega_1 | \ldots | \omega_n] = (-1)^n\big( \sum_{i=1}^n [\omega_1 | \ldots |  d\omega_i| \ldots | \omega_n] +
 \sum_{i=1}^{n-1} [\omega_1 | \ldots |  \omega_i \wedge \omega_{i+1} | \ldots | \omega_n] \big) 
\end{equation} 
and finally project all one-forms on the right-hand side to $\AFbar^1$ via the map $(\ref{projectontoAf})$ and project all two forms (namely, $d \omega_i$ and $\omega_i \wedge \omega_{i+1}$)
onto $\Ao^1_{n-1} \otimes \AFbar^1$  via the decomposition $(\ref{Wedgedecomp})$. Pulling out all factors in $\Ao^1_{n-1}$ to the left gives the required formula for $\nabla$.

The connection $\nabla$ can be promoted to a total connection 
\begin{eqnarray} \label{TotalConnection}
\nabla_T: V(C_n) \longrightarrow \Ao^1_{n} \otimes V(C_n)
\end{eqnarray} 
by setting $\nabla_T= d- \nabla $,  and identifying $\Ao^1_{n-1} \oplus \AFbar_n^1 \cong \Ao^1_{n}$ via the decomposition 
$(\ref{Aosproductdecomp})$. It is straightforward to show that in this context the total connection is flat $(\nabla_T^2=0)$.

Finally, the symbol map is the unique linear map (necessarily a homomorphism)
\begin{equation} \label{SymbolMap}
\Psi: V(C_n) \longrightarrow V(\Mod_{0,n})
\end{equation} 
which  satisfies the equation 
$$ (id \otimes \Psi) \circ \nabla_T =   d \circ \Psi.$$
This can be viewed as  a recursive formula to compute the symbol map $\Psi$  since $\nabla_T$ strictly decreases the length of bar elements. 
Explicitly, it can be rewritten  
$$ \Psi= \int  (id \otimes \Psi) \circ \nabla_T $$
where the total primitive operator $\int$ was defined in \S\ref{sectTotalPrimitives}.

\item  In the other direction, there is the \emph{unshuffle} map which is a homomorphism of graded algebras
\begin{equation}
\Phi:  V(\Mod_{0,n}) \overset{\sim}{\longrightarrow} V(\Mod_{0,n-1}) \otimes V(C_n)
\end{equation} 
which is the inverse of  the map $m(\id \otimes \Psi) : V(\Mod_{0,n-1}) \otimes V(C_n) \rightarrow V(\Mod_{0,n}) $ (which we abusively denote  simply by $\Psi$), where $m$ denotes multiplication.   It can be computed as follows. Denote the  natural map
\begin{eqnarray} r: V(\Mod_{0,n})  & \longrightarrow  & V(C_n)  \nonumber  \\
{[}\omega_1 | \ldots | \omega_r] & \mapsto&  [\overline{\omega}_1|  \ldots  | \overline{\omega}_r] \nonumber 
\end{eqnarray}
given by restriction of iterated integrals  to the fiber  induced by $(\ref{projectontoAf})$
 component-wise on bar elements. Note that the map $\Psi$ has the property that $r\circ \Psi$ is the identity on $V(C_n)$.

Recall the morphism $(\ref{fibration})$ from $\Mod_{0,n}$ to $\Mod_{0,n-1}$ defined in terms of cubical coordinates. 
The projection map $\pi :  \Ao_n \rightarrow \Ao_{n-1}$ implied by the section $\lambda_n$ is given by sending  first $dx_{n-3}$ to zero and then
$x_{n-3}$ to zero. One can see that it is  a homomorphism by inspection of the explicit equations in \S\ref{sub:Arnold-relations}: the product of two elements in ${}^F\!\Ao_n^1$ have no component in $\Ao^2_{n-1}$.
It defines a homomorphism
$$ \pi: V(\Mod_{0,n})\rightarrow V(\Mod_{0,n-1})$$
and one easily verifies that the homomorphism $\Phi$ defined by
$$\Phi( \xi) = (r \otimes  \pi   ) \circ \Delta $$
is an inverse to the symbol map $\Psi$.

Alternatively, we can view $\Mod_{0,n-1}$ as being embedded in $\overline{\Mod}_{0,n}$ by identifying it with the divisor defined by $x_{n-3}=0$.
An element of $V(\Mod_{0,n})$ can be thought of as an iterated integral  along a path from the unit tangential base point at the origin $0$ in cubical
coordinates to  a point $x=(x_1,\ldots, x_{n-3})$. It  is the composition of   a path from  the  unit tangential base point at $0$ to $(x_1,\ldots, x_{n-4})$ in the base $\Mod_{0,n-1}$,
followed by a path in $C_n$ from the unit tangential base point  at $x_{n-3}=0$ to $x$.
Since composition of paths is dual to deconcatenation in $V(\Mod_{0,n})$, this  yields  a geometric interpretation of the above formula for $\Phi$.
\end{enumerate}
 
Thus it is possible, via the symbol and unshuffle maps, to pass back and forth between a representation of an iterated integral on $\Mod_{0,n}$
as a symbol in $V(\Mod_{0,n})$ or a product of words in $V(C_i)$'s.  This gives a precise algorithmic equivalence between the two approaches described in \S\ref{secthypversusmod}. 

\subsection{Representation as functions}
In order to represent elements of $V(\Mod_{0,n})$ as functions (although in principle one never needs to do this) the simplest method is to apply the unshuffle map $\Phi$ defined above, which 
reduces to the problem of representing elements of $V(C_{k})$, for $4\leq k \leq n$ as functions. This is simply the case of computing iterated integrals in a single variable $x_{n-3}$, i.e. hyperlogarithms.
\begin{eqnarray}
V(C_n)  &\longrightarrow & \hbox{Iterated integrals on } C_n  \\ \nonumber 
[ \omega_1 | \ldots | \omega_k]  & \mapsto & \int  \omega_1  \ldots  \omega_k \nonumber
\end{eqnarray}
The iterated integrals on $C_n$ are normalised with respect to the tangential base point ${\partial \over \partial x_{n-3}}$ at $x_{n-3}=0$.
They can be written as polynomials in $\log(x_{n-3})$
and explicit  power series which were studied in  the work of Lappo-Danilevsky \cite{Lap}.  In this way, the iterated use of the unshuffle map reduces the expression of elements
of $V(\Mod_{0,n})$ as functions to a product of hyperlogarithms. These
 are well understood, and can be expressed  in terms of multiple polylogarithms
 $$\mathrm{Li}_{n_1,\ldots, n_r}(x_1,\ldots, x_r) = \sum_{0<k_1<\ldots< k_r} {x_1^{k_1} \ldots x_r^{k_r} \over k_1^{n_1} \ldots k_r^{n_r}}$$
  which can be computed to arbitrary accuracy by standard techniques \cite{Vol}. 
  \subsection{`Mixed' primitives} \label{subMixedPrimitives}
Suppose that we have an element $\xi \in V(\Mod_{0,n})$, and a one form $\omega \in \AF_{n}^1$ which
is only defined on the fiber. The mixed primitive is defined to be 
$$\omega \star \xi := \Psi \big(   \int \omega \,\Phi(\xi)\big)  \qquad \in V(\Mod_{0,n})\ .$$
In other words, $\xi$ is viewed as an element of $V(\Mod_{0,n-1})\otimes V(C_n)$ via the unshuffle map, then multiplied by $1\otimes \omega$
before computing its  primitive $\int$ in $V(C_n)$ (which is simply given by left concatenation of forms in $\AF^1$, as in \S\ref{sectTotalPrimitives}).  Clearly,
the map $\star$ is bilinear  over $\QQ$  and satisfies
\begin{equation}\label{starproperty1} 
 \omega_0 \star \Psi([\omega_1 | \ldots | \omega_k]) = \Psi( [\omega_0 | \ldots | \omega_k])
 \end{equation} 
for all $\omega_i  \in \AFbar_{n}^1$. Furthermore, $\star$ is right-linear over $V(\Mod_{0,n-1})$: 
\begin{equation}\label{starproperty2} 
 \omega \star ( b \shuffle  \xi) = b \shuffle  (\omega \star \xi)
 \end{equation}
for all $b\in V(\Mod_{0,n-1})$, and $\xi \in V(\Mod_{0,n})$, and $\star$ is uniquely determined by $(\ref{starproperty1})$, $(\ref{starproperty2})$ and $(\ref{Vdecomp})$.
Evidently, one does not want to have to compute $\star$ by applying the unshuffling and symbol  maps $\Phi$ and $\Psi$ which would be highly inefficient (and largely redundant).

The approach we have adopted
is more direct. 
Suppose that $\xi =\sum_I  c_I [\omega_{i_1} | \ldots | \omega_{i_m}]$.  As a first approximation to the mixed primitive  $ \omega \star \xi $  take the 
element 
$$ \xi_0=\sum_{I=(i_1,\ldots, i_m)} c_I [\lambda_n(\omega) | \omega_{i_1} | \ldots | \omega_{i_m}]$$
The projection of $\xi_0$ onto $V(C_n)$ coincides with that of $\omega \star \xi$, but $\xi_0$ does not satisfy the integrability condition $(\ref{intcond})$. 
The idea
is to add correction terms $\xi_1, \ldots, \xi_k$ to $\xi_0$  so that the sum $\xi_0 + \ldots + \xi_k = \sum_J c'_J [ \eta_{j_1}| \ldots | \eta_{j_{m+1}}]$  satisfies the  first $k$ integrability equations (with the notation of $(\ref{intcond})$)
$$ \sum_J c'_J [\eta_{i_1} | \ldots | \eta_{j_r} \wedge \eta_{j_{r+1}} | \ldots | \eta_{j_{m+1}}]=0  \qquad \hbox{ for } 1 \leq r \leq k $$
The correction term  $\xi_{k+1}$ is obtained using the quadratic relations $\mu_n$ to expand out each wedge product $\omega_i \wedge \omega_j$ in the $k+1$th integrability 
equation, applied to $\xi_0+\ldots+ \xi_k$.
The mixed primitive $\omega \star \xi$ is equal to the sum $\xi_0+ \ldots + \xi_{m}$ if $\xi$ is of length $m$.
The precise details are described below.

\subsection{Feasibility and orders of magnitude}

By iterating theorem $(\ref{thmVdecomp})$  one obtains a formula for the dimension of all symbols on $\Mod_{0,n+3}$ in weight $N$:
\begin{equation} \label{dimformula}
\sum_{N\geq 0}\, (\dim_{\QQ} V(\Mod_{0,n+3})_{N}) t^N = {1 \over (1-2t) (1-3t) \ldots (1-(n+1)t)}
\end{equation}  
This gives a coarse upper bound for the possible size of expressions which can occur during the integration process.
At the initial step of integration, the integrand is of weight $0$ on a moduli space of high dimension $\Mod_{0,n+3}$, and at the final
step, the integrand is of high weight on a moduli space of low dimension $\Mod_{0,4}$. The dimensions $(\ref{dimformula})$ peak
somewhere in the middle of the computation.  For example, for (the maximal weight part) of a period integral $(\ref{introM0nint})$ in five variables,
one works in a sequence of vector spaces of dimension $20,125,285, 211,32$ (these are the dimensions of the spaces  of functions after taking each primitive and before taking each limit).

In the case of Feynman diagrams, one can estimate in advance  (using the linear reduction algorithm) the number of marked points $n$ which will be required at each step of the integration to get
a bound on the size of the computation. In practice, it seems that the limit of what is reasonable with current levels of computing power should be adequate to 
reach the `non-polylogarithmic' boundary where  amplitudes which are not periods of mixed Tate motives first start to appear.

\section{Computing on the moduli space}\label{sec:Computing-on-the}

In this section we spell out the details of the above algorithms and
present them in a version which is ready for  implementation on
a computer. As a proof of  concept we implemented these algorithms
in a  Maple-based computer program. With this program we computed
all examples below and all applications of section \ref{sec:Applications}.

For notational convenience let $m=n-3$ denote the number of cubical coordinates $x_i$ on $\Mod_{0,n}$. As bases for $\Ao^1_{n}$, $\AFbar^1_n$ and $\AF^1_n$ we choose the sets of closed 1-forms
\begin{eqnarray*}
\Omega_{m} & = & \left\{ \frac{dx_{1}}{x_{1}},\,...,\,\frac{dx_{m}}{x_{m}},\,\frac{d\left(\prod_{a\leq i\leq b}x_{i}\right)}
{\prod_{a\leq i\leq b}x_{i}-1}\textrm{ for }1\leq a\leq b\leq m\right\} ,\\
\bar{\Omega}^F_{m} & = & \left\{ \frac{dx_{m}}{x_{m}},\,\frac{\left(\prod_{a\leq i\leq m-1}x_{i}\right)dx_{m}}
{\prod_{a\leq i\leq m}x_{i}-1}\textrm{ for }1\leq a\leq m\right\} ,\\
\Omega_{m}^{F} & = & \left\{ \frac{dx_{m}}{x_{m}},\,\frac{d\left(\prod_{a\leq i\leq m}x_{i}\right)}{\prod_{a\leq i\leq m}x_{i}-1}
\textrm{ for }1\leq a\leq m\right\} ,\\
\end{eqnarray*}
respectively. The isomorphism $\bar{\AF_n}\overset{\lambda_{n}}{\cong}\AF_n \subseteq \Ao_n$ of $(\ref{lambdadefn})$ is defined explicitly by 
\begin{eqnarray} \label{lambdaexplicit} 
\lambda_{n}\frac{dx_{m}}{x_{m}} & = & \frac{dx_{m}}{x_{m}},\\
\lambda_{n}\frac{\left(\prod_{a\leq i\leq m-1}x_{i}\right)dx_{m}}{\prod_{a\leq i\leq m}x_{i}-1} & = & 
\frac{d\left(\prod_{a\leq i\leq m}x_{i}\right)}{\prod_{a\leq i\leq m}x_{i}-1}\textrm{ for }1\leq a \leq m. \nonumber 
\end{eqnarray}

According to these chosen bases, we refer to the vector-spaces $V(C_{n}),\, V(\mathcal{M}_{0,n})$
by $V(\Omega^F_{m}),\, V(\Omega_{m})$ respectively. Iterated integrals are written as linear combinations of words $[\omega_{1}|...|\omega_{k}]$, whose letters are 
1-forms in these sets.  Note that $\Omega_m $ is the disjoint union of $\Omega_{m-1}$ and $\Omega_{m}^F$.

\subsection{Arnold relations}\label{sub:Arnold-relations}

With the above choices, the Arnold relations of $(\ref{Wedgedecomp})$  read explicitly:
\begin{eqnarray*}
\frac{dx_{m}}{x_{m}}\wedge\frac{d\left(x_{i}...x_{m}\right)}{x_{i}...x_{m}-1} & = & -\sum_{k=i}^{m-1}\frac{dx_{k}}{x_{k}}\wedge\frac{d\left(x_{i}...x_{m}\right)}{x_{i}...x_{m}-1},\\
\frac{d\left(x_{j}...x_{m}\right)}{x_{j}...x_{m}-1}\wedge\frac{d\left(x_{i}...x_{m}\right)}{x_{i}...x_{m}-1} & = & \frac{d\left(x_{i}...x_{j-1}\right)}{x_{i}...x_{j-1}-1}\wedge\left(\frac{d\left(x_{i}...x_{m}\right)}{x_{i}...x_{m}-1}-\frac{d\left(x_{j}...x_{m}\right)}{x_{j}...x_{m}-1}\right)-\sum_{k=i}^{j-1}\frac{dx_{k}}{x_{k}}\wedge\frac{d\left(x_{i}...x_{m}\right)}{x_{i}...x_{m}-1}
\end{eqnarray*}
for $1\leq i\leq j\leq m.$ For the implementation on a computer,
it is efficient to generate these equations to a desired number
of variables once and for all, and to store them as a look-up table since they are used very frequently by the algorithms below.

The splitting of theorem \ref{thmVdecomp} is realised by a certain application
of the Arnold relations. We define an auxiliary map $\rho_{i}$ by the following operations. For a word $\xi=[\omega_{1}|...|\omega_{k}]$ with letters in 
$\bar{\Omega}^F_{m}$ and some $1\leq i<k$ we consider 
the neighbouring letters $\omega_i | \omega_{i+1}$ and consider the wedge-product of their images in $\Omega^F_{m}$. 
By the corresponding Arnold relation, we express this product as a $\mathbb{Q}$-linear combination 
of wedge-products, with one factor in the base $\Omega_{m-1}$ and one in the fiber $\Omega^F_{m}$. We replace the letters $\omega_i | \omega_{i+1}$ in $\xi$ by the 
factor in $\Omega^F_{m}$ and pull the base-term in $\Omega_{m-1}$ and rational pre-factors out of the word. In summary, this defines the 
 auxiliary map 
\[
\rho_{i}:\, V\left(\bar{\Omega}^F_{m}\right)\rightarrow\Omega_{m-1}\otimes V\left(\bar{\Omega}^F_{m}\right)
\]
 by 
\[
\rho_{i}\left[a_{1}|...|a_{k}\right]=\sum_{j}c_{j}\eta_{j}\otimes\left[a_{1}|...|a_{i-1}| \overline{\alpha}_{j}|a_{i+2}|...|a_{k}\right]
\]
where  $\eta_{j}\in\Omega_{m-1},\,\alpha_{j}\in\Omega_{m}^{F},\, c_{j}\in\mathbb{Q}$
are determined by the Arnold relation 
\[
\lambda_{n}a_{i}\wedge\lambda_{n}a_{i+1}=\sum_{j}c_{j}\eta_{j}\wedge\alpha_{j}.
\]
Note that these are the same operations as in our definition of the Gauss-Manin connection $\nabla$ above, which we obtain by summing the $\rho_i$ over $i$. 
This is because the first sum on the right-hand side of $(\ref{integrability})$ vanishes in our
set-up, as all our 1-forms are closed, and the operations on the terms of the second sum correspond to the definition of $\rho_i$. 

\begin{example} \label{exampleArnoldn=5}
For $n=5, m=2$ we have the Arnold relations
\begin{eqnarray*}
\frac{x_{1}dx_{2}+x_{2}dx_{1}}{x_{1}x_{2}-1}\wedge\frac{dx_{2}}{x_{2}} & = & \frac{dx_{1}}{x_{1}}\wedge\frac{x_{1}dx_{2}+x_{2}dx_{1}}{x_{1}x_{2}-1},\\
\frac{x_{1}dx_{2}+x_{2}dx_{1}}{x_{1}x_{2}-1}\wedge\frac{dx_{2}}{x_{2}-1} & = & \left(\frac{dx_{1}}{x_{1}}-\frac{dx_{1}}{x_{1}-1}\right)\wedge\frac{x_{1}dx_{2}+x_{2}dx_{1}}{x_{1}x_{2}-1}+\frac{dx_{1}}{x_{1}-1}\wedge\frac{dx_{2}}{x_{2}-1}.
\end{eqnarray*}

For the words $\kappa=\left[\frac{x_{1}dx_{2}}{x_{1}x_{2}-1}|\frac{dx_{2}}{x_{2}-1}\right],$
$\xi=\left[\frac{x_{1}dx_{2}}{x_{1}x_{2}-1}|\frac{dx_{2}}{x_{2}}|\frac{dx_{2}}{x_{2}-1}\right]$
in $V(\bar{\Omega}^F_{2})$ we compute 
\begin{eqnarray*}
\rho_{1}\kappa & = & \left[\frac{dx_{1}}{x_{1}}\right]\otimes\left[\frac{x_{1}dx_{2}}{x_{1}x_{2}-1}\right]
-\left[\frac{dx_{1}}{x_{1}-1}\right]\otimes\left[\frac{x_{1}dx_{2}}{x_{1}x_{2}-1}\right]+\left[\frac{dx_{1}}{x_{1}-1}\right]
\otimes\left[\frac{dx_{2}}{x_{2}-1}\right],\\
\rho_{1}\xi & = & \left[\frac{dx_{1}}{x_{1}}\right]\otimes\left[\frac{x_{1}dx_{2}}{x_{1}x_{2}-1}|\frac{dx_{2}}{x_{2}-1}\right],\\
\rho_{2}\xi & = & 0.
\end{eqnarray*}

\end{example}

\subsection{The symbol map} \label{sub:The-symbol-map}

Both the total connection and the symbol map can be computed conveniently
by use of the maps $\rho_{i}$. The total connection (see $(\ref{TotalConnection})$) is computed as 

\[
\nabla_{T}\left[a_{1}|...|a_{k}\right]=d\left[a_{1}|...|a_{k}\right]-\sum_{1\leq i<k}\rho_{i}\left[a_{1}|...|a_{k}\right]
\]
where (by $(\ref{BdRdifferential})$)
\[
d\left[a_{1}|...|a_{k}\right]=a_{1}\otimes\left[a_{2}|...|a_{k}\right].
\]
The symbol map $\Psi$ (see $(\ref{SymbolMap})$) is applied to a word in $V(\bar{\Omega}^F_{m})$
by the recursive algorithm
\begin{eqnarray}
\Psi\left(\left[a_{i}\right]\right) & = & \left[\lambda_{n}\left(a_{i}\right)\right],\nonumber \\
\Psi\left(\left[a_{i_{1}}|a_{i_{2}}|...|a_{i_{k}}\right]\right) & = & \lambda_{n}\left(a_{i_{1}}\right)\sqcup\Psi\left(\left[a_{i_{2}}|...|a_{i_{k}}\right]\right)-\sum_{1\leq i<k}\sqcup\left(\left(\textrm{id}\otimes\Psi\right)\rho_{i}\left[a_{i_{1}}|...|a_{i_{k}}\right]\right),\;1<k,\label{eq:symbol map-1}
\end{eqnarray}
where $\xi_1\sqcup \xi_2\equiv\sqcup(\xi_1\otimes \xi_2)$ denotes the concatenation of two words $\xi_1,\, \xi_2.$ Note that on the
right hand side of $(\ref{eq:symbol map-1})$ the map $\Psi$ acts
on words of length $k-1.$
\begin{example}
Making use of the relations derived in  example \ref{exampleArnoldn=5},
we compute 
\begin{eqnarray*}
\Psi\left(\left[\frac{x_{1}dx_{2}}{x_{1}x_{2}-1}|\frac{dx_{2}}{x_{2}}|\frac{dx_{2}}{x_{2}-1}\right]\right) & = & \left[\frac{x_{1}dx_{2}+x_{2}dx_{1}}{x_{1}x_{2}-1}|\frac{dx_{2}}{x_{2}}|\frac{dx_{2}}{x_{2}-1}\right]-\left[\frac{dx_{1}}{x_{1}}\right]\sqcup\Psi\left(\left[\frac{x_{1}dx_{2}}{x_{1}x_{2}-1}|\frac{dx_{2}}{x_{2}-1}\right]\right)\\
 & = & \left[\frac{x_{1}dx_{2}+x_{2}dx_{1}}{x_{1}x_{2}-1}|\frac{dx_{2}}{x_{2}}|\frac{dx_{2}}{x_{2}-1}\right]-\left[\frac{dx_{1}}{x_{1}}|\frac{x_{1}dx_{2}+x_{2}dx_{1}}{x_{1}x_{2}-1}|\frac{dx_{2}}{x_{2}-1}\right]\\
 &  & -\left[\frac{dx_{1}}{x_{1}}|\frac{dx_{1}}{x_{1}-1}|\frac{x_{1}dx_{2}+x_{2}dx_{1}}{x_{1}x_{2}-1}\right]+\left[\frac{dx_{1}}{x_{1}}|\frac{dx_{1}}{x_{1}-1}|\frac{dx_{2}}{x_{2}-1}\right]\\
 &  & +\left[\frac{dx_{1}}{x_{1}}|\frac{dx_{1}}{x_{1}}|\frac{x_{1}dx_{2}+x_{2}dx_{1}}{x_{1}x_{2}-1}\right].
\end{eqnarray*}

\end{example}
The map $\Psi$ is defined such that for any $\xi\in V\left(\bar{\Omega}^F_{m}\right)$
we have $D\Psi\left(\xi\right)=0$ and therefore $\Psi\left(\xi\right)\in V\left(\Omega_{m}\right).$
The vector space $V\left(\Omega_{m}\right)$ is generated, over $V(\Omega_{m-1})$, by the image
of $V\left(\bar{\Omega}^F_{m}\right)$ under $\Psi.$ We furthermore
note the property 
\[
\Psi\left(\xi_{1}\shuffle\xi_{2}\right)=\Psi(\xi_{1})\shuffle\Psi(\xi_{2})
\]
for any $\xi_{1},\,\xi_{2}\in V\left(\bar{\Omega}^F_{n}\right).$ 

A slightly different algorithm for $\Psi$ in terms of differentiation under the integral was already given in \cite{Bog1}. For 
related constructions, also see references \cite{Duh1, Gon1, Gon2}.
In section \ref{sec:Feynman-type-integrals} we will make use of $\Psi$
as a part of a procedure to map hyperlogarithms in Schwinger parameters to multiple polylogarithms
of cubical variables. We expect the map $\Psi$ also to be useful in different contexts such as \cite{Druetal}.

\subsection{Primitives}\label{sub:Primitives}

Let $\omega\in\bar{\Omega}^F_m$ and let $\xi=\sum_{I}c_{I}\left[\omega_{i_{1}}|...|\omega_{i_{k}}\right]$ be an iterated integral in $V(\Omega_{m})$. 
In subsection \ref{subMixedPrimitives}, we discussed the strategy of building up the mixed primitive $\omega\star\xi$ by naive left-concatenation of the 
form $\omega$ to the word $\xi$, yielding
\begin{equation}  \label{xi0ofmixedprimitives}
\sum_{I}c_{I}\left[\lambda_{n}(\omega)|\omega_{i_{1}}|...|\omega_{i_{k}}\right],
\end{equation}
 and the addition of correction terms 
until the resulting combination satisfies the integrability condition of $(\ref{intcond})$.
For the explicit computation of the correction terms, let us introduce some auxiliary notation. For all $0\leq i < k$
let $C_{i}\left(\Omega_{m}\right)_k= \Omega_{m-1}^{\otimes i} \otimes \Omega_m^F \otimes  \Omega_m^{\otimes(k-i-1)}$ be the $\mathbb{Q}$-vector space
of words of length $k$ with letters in $\Omega_{m}$, whose  first $i$ letters,  counted from the left, are in  the base $\Omega_{m-1}$, and whose $(i+1)^{\mathrm{th}}$ letter is in the fiber
$\Omega_m^F$.
The members of these auxiliary sets of words do not necessarily
stand for homotopy invariant iterated integrals. 
We define the auxiliary maps 
\[
\star_{i}:\, C_{i-1}\left(\Omega_{m}\right)_k\rightarrow C_{i}\left(\Omega_{m}\right)_k
\]
 for $i<k$ by the following recipe
\begin{eqnarray}
\star_{i}[a_{1}|...|a_{i-1}|a_{i}|a_{i+1}|...|a_{k}] & = & [a_{1}|...|a_{i-1}|a_{i+1}|a_{i}|...|a_{k}]\qquad \textrm{ if  } a_{i+1}\in\Omega_{m-1} , \nonumber \\
\star_{i}[a_{1}|...|a_{i-1}|a_{i}|a_{i+1}|...|a_{k}] & = & -\sum_{j}c_{j}[a_{1}|...|a_{i-1}|\eta_{j}|\alpha_{j}|a_{i+2}|...|a_{k}]\qquad \textrm{ if }a_{i+1}\in\Omega_{m}^{F},\, \label{eq:star i}
\end{eqnarray}
where the forms $\eta_{j}\in\Omega_{m-1},\,\alpha_{j}\in\Omega_{m}^{F}$ and constants $\, c_{j}\in\mathbb{Q}$
are determined by an Arnold relation 
\[
a_{i}\wedge a_{i+1}=\sum_{j}c_{j}\eta_{j}\wedge\alpha_{j}.
\]
Note that indeed, in each word on the right-hand side of $(\ref{eq:star i})$
the 1-forms in the first $i$ positions are in $\Omega_{m-1}$ and the form in  the $(i+1)$-th position is  in $\Omega_{m}^{F}$.
This procedure can be iterated. Since $\ref{xi0ofmixedprimitives}$ lies in $C_{0}\left(\Omega_{m}\right)_ {k+1}$,
we repeatedly apply  $\star_{\bullet}$ to obtain the following formula for the mixed primitive 
\begin{eqnarray} \label{eq:Primitive}
\omega\star[a_{1}|...|a_{k}]=(1+\star_{1}+\star_{2}\star_{1}+...+\star_{k}...\star_{1})[\lambda_{m}(\omega)|a_{1}|...|a_{k}].
\end{eqnarray}

The construction satisfies the relations  $(\ref{starproperty1})$ and $(\ref{starproperty2})$.
\begin{example}
We consider the 1-form $\omega = \frac{dx_2}{x_2}$, the iterated integral 
\begin{eqnarray*}
\xi  =  \Psi \left(\left[ \frac{x_1 d(x_2)}{x_1 x_2 -1} | \frac{dx_2}{x_2}  \right] \right) =  \left[ \frac{d(x_1 x_2)}{x_1 x_2 -1} | \frac{dx_2}{x_2}  \right]  - \left[ \frac{dx_1}{x_1} | \frac{d(x_1 x_2)}{x_1 x_2 -1}   \right],
\end{eqnarray*}
and the concatenation 
\begin{eqnarray*}
\xi_0 & = & \lambda_2 (\omega) \sqcup \xi =  \left[ \frac{dx_2}{x_2} | \frac{d(x_1 x_2)}{x_1 x_2 -1} | \frac{dx_2}{x_2}  \right] - \left[ \frac{dx_2}{x_2} | \frac{dx_1}{x_1} | \frac{d(x_1 x_2)}{x_1 x_2 -1} \right].
\end{eqnarray*}
Following $(\ref{eq:Primitive})$, we compute the primitive
\begin{eqnarray*}
\omega \star \xi = \xi_0 + \xi_1 + \xi_2
\end{eqnarray*}
where $\xi_1 = \star_1 \xi_0$ and $\xi_2 = \star_2 \star_1 \xi_0$. We obtain 
\begin{eqnarray*}
 \\
\xi_1  & = & \left[ \frac{dx_1}{x_1} | \frac{d(x_1 x_2)}{x_1 x_2 -1} | \frac{dx_2}{x_2} \right] - \left[ \frac{dx_1}{x_1} | \frac{dx_2}{x_2} | \frac{d(x_1 x_2)}{x_1 x_2 -1} \right], \\
\xi_2 & = & -2\left[ \frac{dx_1}{x_1} | \frac{dx_1}{x_1} | \frac{d(x_1 x_2)}{x_1 x_2 -1} \right]
\end{eqnarray*}
by use of the Arnold relations given in the example of section \ref{sub:Arnold-relations}.
\end{example}

\subsection{Limits} \label{sub:Limits} We consider limits at $x_{l}=u$, $l\in\{1,\,...,\, m\}$ where $u\in\{0,\,1\}.$
By definition, any $\xi\in V\left(\Omega_{m}\right)$ vanishes along 
$x_{l}=0.$ Limits at $0$ and $1$ are computed as follows.

As in the previous sections, let $\mathcal{Z}$ be the $\mathbb{Q}$-vector space of multiple
zeta values. It was shown in \cite{Bro2} that for any $\xi\in V\left(\Omega_{m}\right)$
the limits $\lim_{x_{l}\rightarrow 1}\xi$ are $\mathcal{Z}$-linear
combinations of elements of $V\left(\Omega_{m-1}\right) $ 
(after a possible renumbering of the cubical coordinates: $(x_{l+1},\ldots, x_m) \mapsto (x_l,\ldots,  x_{m-1}$).)
Our algorithm for the computation of limits proceeds in two steps: 
\begin{itemize}
\item Expand the function $\xi$ at $x_{l}=u$ as a polynomial in  $\log(x_{l}-u)$, whose coefficients are power series in $x_{l}-u$, and 
\item Evaluate the constant term (coefficient of  $\log(x_l-u)^0$)  at $x_{l}=u.$ 
\end{itemize}
The series expansion is the non-trivial part in this computation while the evaluation of the series is immediate. Let $\textrm{Exp}_{x_{l}=u}\xi(x_{l})$
denote the expansion of the function $\xi(x_{l})$ at $x_{l}=u.$
We compute the expansion recursively as 
\begin{equation} \label{eq:Expansion}
\textrm{Exp}_{x_{l}=u}\xi(x_{l})=\textrm{Reg}_{x_{l}=u}\xi(x_{l})+\int dx'_{l}\,\textrm{Exp}_{x'_{l}=u}\frac{\partial}{\partial x'_l}\xi(x'_{l}).
\end{equation}
where the integral on the right is the regularised integral from the tangential base point ${\partial \over \partial x_l}$ at $x_l=u$ to $x_l$, 
or equivalently, is an indefinite integral in $x_l$ whose constant of integration is fixed by declaring that its regularised limit at $x_l=u$ vanishes.
Note that if $\xi(x_{l})$ is a linear combination of  words of  length $k$, then in the integrand on the
right-hand side of $(\ref{eq:Expansion})$, $\textrm{Exp}_{x'_{l}=u}$
is computed  on words of length $k-1$. Rational prefactors are trivially expanded
as power series in $x_l=u$ also. The notation $\textrm{Reg}_{x_{l}=u}\xi(x_{l})$ stands for
the operation of taking the regularised limit of $\xi$ at $x_{l}=u.$
For $u=0$ we define $\textrm{Reg}_{x_{l}=0}$ to be the identity-map on terms of weight $0$ and 
\[
\textrm{Reg}_{x_{l}=0}\xi(x_{l})=0
\]
for $\xi(x_{l})$ with all terms of weight greater than $0$. 
For $u=1$ regularised limits are defined and computed in the remainder
of this subsection.

Let us start by computing regularised limits of iterated integrals
in only one variable and then extend to the $n$-variable case. We
consider $\Omega_{1}=\left\{ \frac{dx_{1}}{x_{1}},\,\frac{dx_{1}}{x_{1}-1}\right\} $
and for $\xi\in V\left(\Omega_{1}\right)$ we use a simplified notation
where in each word we symbolically replace $\frac{dx_{1}}{x_{1}}\rightarrow0$
and $\frac{dx_{1}}{x_{1}-1}\rightarrow1$ and multiply the word with
$(-1)^{s}$ where $s$ is the number of 1-forms $\frac{dx_{1}}{x_{1}-1}.$
Following \cite{Bro1} we define the map 
\[
\textrm{Reg}_{x_{1}=1}:\, V\left(\Omega_{1}\right)\rightarrow\mathcal{Z}
\]
by the following relations for different cases of words $\xi=\left[a_{1}|...|a_{k}\right],\, a_{i}\in\{0,\,1\},\, i=1,\,...,\, k$:
\begin{itemize}
\item Case 1: If all letters are equal, $a_{1}=a_{2}=...=a_{k}$, we have
\[
\textrm{Reg}_{x_{1}=1}\left[a_{1}|...|a_{k}\right]=0.
\]

\item Case 2: If the word begins with 0 and ends with 1 (from left to right), we have 
\[
\textrm{Reg}_{x_{1}=1}[\underbrace{0|...|0|1|}_{n_{r}}...|1|\underbrace{0|...|0|1}_{n_{1}}]=\zeta(n_1,\,...,\, n_{r})\textrm{ for }n_{r}\geq2,\, n_{i}\geq1,\, n_{1}+...+n_{r}=k.
\]

\item Case 3: If the word begins in  1 we apply the relation
\[
\textrm{Reg}_{x_{1}=1}\left[a_{1}|...|a_{k}\right]=\textrm{Reg}_{x_{1}=1}\left[1-a_{k}|...|1-a_{1}\right]
\]
which is also true in all other cases. 
\item Case 4: If the word ends with 0 we use the relation
\[
\textrm{Reg}_{x_{1}=1}[\underbrace{0|...|0|1|}_{n_{1}}...|1|\underbrace{0|...|0|1}_{n_{r}}|\underbrace{0|...|0}_{q}]=
\]
\begin{equation}
(-1)^{q}\sum_{i_{1}+...+i_{r}=q}\left(\begin{array}{c}
n_{1}+i_{1}-1\\
i_{1}
\end{array}\right)...\left(\begin{array}{c}
n_{r}+i_{r}-1\\
i_{r}
\end{array}\right)\textrm{Reg}_{x_{1}=1}[\underbrace{0|...|0|1|}_{n_{1}+i_{1}}...|1|\underbrace{0|...|0|1}_{n_{r}+i_{r}}],\label{eq:case 4}
\end{equation}
where $q,\, n_{1},\,...,\, n_{r}\geq1.$
\end{itemize}
By these relations, implementing the well-known shuffle-regularization, the regularized value of any $\xi\in V\left(\Omega_{1}\right)$
can be expressed as a $\mathbb{Q}$-linear combination of expressions
as in case 2, which are  multiple zeta values.

\begin{example}
We consider $\xi=\left[\frac{dx_{1}}{x_{1}-1}|\frac{dx_{1}}{x_{1}}|\frac{dx_{1}}{x_{1}}\right]$
which in short-hand notation reads $\xi=-[1|0|0]$ and falls into
the above case 4. By $(\ref{eq:case 4})$ we have $\textrm{Reg}_{x_{1}=1}(-[1|0|0])=\textrm{Reg}_{x_{1}=1}(-[0|0|1])$
and obtain by case 2: 
\[
\textrm{Reg}_{x_{1}=1}\xi=-\zeta(3).
\]

\end{example}
Now we extend the definition of regularized limits to $V\left(\Omega_{m}\right).$
Let us  first define the auxiliary restriction maps
\begin{eqnarray*}
R_{x_{l}}:\, V\left(\Omega_{m}\right) & \rightarrow & V\left(\Omega_{1}\right)
\end{eqnarray*}
by 
\begin{equation}
R_{x_{l}}\xi=\xi|_{dx_{i}=0,\, x_{i}=0\textrm{ for all }i\in\{1,\,...,\, m\},\, i\neq l}\label{eq:restriction R}
\end{equation}
and 
\[
L_{x_{l}}:\, V\left(\Omega_{m}\right)\rightarrow V\left(\Omega_{m-1}\right)
\]
by 
\begin{equation}
L_{x_{l}}\xi=\xi|_{dx_{l}=0,\, x_{l}=1}.\label{eq:restriction L}
\end{equation}
and relabelling cubical coordinates as mentioned above. Note that the map $R_{x_{l}}$ is the projection onto words all of whose 1-forms are $\frac{dx_l}{x_l}$ and $\frac{dx_l}{x_l-1}$.

These maps play a similar role as the restrictions $E_\gamma$ in section \ref{sectbardeRham}. The map $R_{x_{l}}$ restricts the iterated integral to the 
straight line from the origin to $1_l$ (called $\gamma_2$ in section \ref{sectbardeRham}) and $L_{x_{l}}$ restricts to the divisor of $\overline{\Mod}_{0,n}$ defined by  $x_l=1$ (in which $\gamma_1$ of
section \ref{sectbardeRham} lives). 
According to $(\ref{convolution})$, we take the deconcatenation co-product $\Delta$ of $\xi\in V\left(\Omega_{m}\right)$
and apply $L_{x_{l}}$ and $R_{x_{l}}$ to the left and right part
of the tensor product respectively. The right-hand side of the tensor
product is then in $V\left(\Omega_{1}\right)$ and we apply the above
map of regularized values to this part. In summary, we extend the
definition of regularized values to 
\begin{equation*}
\textrm{Reg}_{x_{l}=1}:\, V\left(\Omega_{m}\right)\rightarrow\mathcal{Z}\otimes V\left(\Omega_{m-1}\right)
\end{equation*}
by 
\begin{equation} \label{eq:Reg}
\textrm{Reg}_{x_{l}=1}\xi=m\left(L_{x_{l}}\otimes \textrm{Reg}_{x_{l}=1} R_{x_{l}}\right)\circ \Delta\xi.
\end{equation}

This completes our algorithm for computing limits of $\xi\in V\left(\Omega_{m}\right)$
at $x_{l}=0,\,1.$
\begin{example}
We consider the iterated integral 
\begin{eqnarray*}
\xi & = & \Psi\left(\left[\frac{x_{1}dx_{2}}{x_{1}x_{2}-1}|\frac{dx_{2}}{x_{2}}|\frac{dx_{2}}{x_{2}-1}\right]\right)\\
 & = & \left[\frac{d(x_{1}x_{2})}{x_{1}x_{2}-1}|\frac{dx_{2}}{x_{2}}|\frac{dx_{2}}{x_{2}-1}\right]-\left[\frac{dx_{1}}{x_{1}}|\frac{d(x_{1}x_{2})}{x_{1}x_{2}-1}|\frac{dx_{2}}{x_{2}-1}\right]-\left[\frac{dx_{1}}{x_{1}}|\frac{dx_{1}}{x_{1}-1}|\frac{d(x_{1}x_{2})}{x_{1}x_{2}-1}\right]\\
 &  & +\left[\frac{dx_{1}}{x_{1}}|\frac{dx_{1}}{x_{1}-1}|\frac{dx_{2}}{x_{2}-1}\right]+\left[\frac{dx_{1}}{x_{1}}|\frac{dx_{1}}{x_{1}}|\frac{d(x_{1}x_{2})}{x_{1}x_{2}-1}\right]\\
 & \in & V\left(\Omega_{2}\right).
\end{eqnarray*}
In this case, the only contibutions to the limit at $x_2=1$ are given by the term $\textrm{Reg}_{x_2 =1} \xi (x_2)$ of $(\ref{eq:Expansion})$, which we compute 
by use of $(\ref{eq:Reg})$. The coproduct of $\xi$ involves 20 terms, most of which vanish after applying $L_{x_2}$ to the left and $R_{x_2}$ to the right part. 
From the non-vanishing terms we obtain
\begin{eqnarray*}
\textrm{Reg}_{x_2 =1} \xi (x_2) & = & m \left( \left[ \frac{dx_1}{x_1 -1} \right] \otimes \textrm{Reg}_{x_2 =1} \left[ \frac{dx_2}{x_2} | \frac{dx_2}{x_2 -1} \right]
-  \left[  \frac{dx_1}{x_1} | \frac{dx_1}{x_1 -1} \right] \otimes \textrm{Reg}_{x_2 =1} \left[ \frac{dx_2}{x_2 -1} \right]  \right. \\
& &  + \left[ \frac{dx_1}{x_1} | \frac{dx_1}{x_1} | \frac{dx_1}{x_1-1} \right] \otimes 1 - \left[ \frac{dx_1}{x_1} | \frac{dx_1}{x_1-1} | \frac{dx_1}{x_1-1} \right] \otimes 1 \\
& & \left.+\left[ \frac{dx_1}{x_1} | \frac{dx_1}{x_1-1} \right] \otimes \textrm{Reg}_{x_2 =1} \left[ \frac{dx_2}{x_2 -1} \right] \right).
\end{eqnarray*}
Due to 
\begin{eqnarray*}
 \textrm{Reg}_{x_2 =1} \left[ \frac{dx_2}{x_2 -1} \right]=0 \textrm{ \ \ and \ \ } \textrm{Reg}_{x_2 =1} \left[ \frac{dx_2}{x_2} | \frac{dx_2}{x_2 -1} \right] = -\zeta(2)
\end{eqnarray*}
or by cancellation of the second and fifth terms, we obtain the limit
\begin{eqnarray*}
\lim_{x_{2}\rightarrow1}\xi=\left[\frac{dx_{1}}{x_{1}}|\frac{dx_{1}}{x_{1}}|\frac{dx_{1}}{x_{1}-1}\right]-\left[\frac{dx_{1}}{x_{1}}|\frac{dx_{1}}{x_{1}-1}|\frac{dx_{1}}{x_{1}-1}\right]-\zeta(2)\left[\frac{dx_{1}}{x_{1}-1}\right].
\end{eqnarray*}

\end{example}

\section{Feynman-type integrals} \label{sec:Feynman-type-integrals}

In this section we consider finite integrals derived from (linearly reducible, unramified) Feynman
integrals. We present an algorithm to map such integrals to hyperlogarithms
in cubical variables (corresponding to the morphism $X\rightarrow \overline{\Mod}_{0,n+1}$ in the diagram \ref{Square}). The integration over one chosen Schwinger 
parameter maps to the integration over one cubical variable. Then this integration 
can be computed by the algorithms of section \ref{sec:Computing-on-the}.
After the integration, as a preparation for the integration over a
next Schwinger parameter, we map back to iterated integrals in Schwinger parameters.

\subsection{Schwinger  parameters}

In dimensional regularization, scalar Feynman integrals of Feynman
graphs $G$ with $N$ edges and loop-number $L,$ can be written in
the Feynman parametric form 
\[
I_{G}(D)=\frac{\Gamma(\nu-LD/2)}{\prod_{j=1}^{n}\Gamma(\nu_{j})}\int_{\alpha_{j}\geq0}\delta\left(1-\alpha_{N}\right)\left(\prod_{j=1}^{N}d\alpha_{j}\alpha_{j}^{\nu_{j}-1}\right)\frac{\mathcal{U}_{G}^{\nu-(L+1)D/2}}{\mathcal{F}_{G}{}^{\nu-LD/2}},
\]
where $\nu=\sum_{i=1}^{N}\nu_{i}$ is the sum of powers of Feynman
propagators and $D\in\mathbb{C}$. We refer to the variables $\alpha_{1},\,...,\,\alpha_{N}$
as Schwinger parameters and the above integration is over the positive
range of each of these variables. The functions $\mathcal{U}_{G}$
and $\mathcal{F}_{G}$ are the first and second Symanzik polynomials respectively. They are 
polynomials in the Schwinger parameters and $\mathcal{F}_{G}$
is furthermore a polynomial of kinematical invariants, which are quadratic
functions of particle masses and external momenta of $G.$ For more
details we refer to \cite{Itz, Nak, Bog2}. 

Assume that we want to compute $I_{G}(2n)$ for some $n\in\mathbb{N}.$
There are different scenarios in which our algorithms may be useful.
In the simplest case, the integral $I_{G}(2n)$ is finite and we may
attempt to compute it without further preparative steps. If $I_{G}(2n)$
is divergent there may be a $n\neq m\in\mathbb{N}$ such that $I_{G}(2m)$
is finite and the method of \cite{Tar1, Tar2} may provide useful
relations between $I_{G}(2n)$ and $I_{G}(2m).$ These relations however
may involve further integrals to be computed. The method of \cite{Bro3}
allows for a subtraction of UV divergent contributions by a renormalization
procedure on the level of the integrand. Alternatively, for a possibly
UV and IR divergent integral, we may attempt to expand $I_{G}$ as
\[
I_{G}=\sum_{j=-2L}^{\infty}c_{j}\epsilon^{j}
\]
where $\epsilon=(2n-D)/2$ and the $c_{j}$ are finite integrals.
In principle such an expansion can be computed by sector decomposition \cite{Bin},
however in this case, a use of our algorithms may be prohibited by
the type of polynomials appearing in the integrands of the resulting $c_{j}.$
Recently, an alternative approach, where the latter polynomials are
given by Symanzik polynomials of $G$ and its minors was suggested
in \cite{Pan2}.

Let us assume that these or alternative methods have led us to an
integral over the positive range of Schwinger  parameters where the integrand
is of the form 
\begin{equation} \label{hypIntegrand}
f(\alpha_{1},\,...,\,\alpha_{N})=\frac{\prod_{Q_{i}\in\mathcal{Q}}Q_{i}^{\delta_{i}}\textrm{ hyperlogarithms}({P_{i}})}{\prod_{P_{i}\in\mathcal{P}}P_{i}^{\beta_{i}}}
\end{equation}
where all $\delta_{i},\,\beta_{i}\in\mathbb{N}_{0}$ and where $\mathcal{P}=\{P_{1},\,...,\, P_{r}\}$
and $\mathcal{Q}=\{Q_{1},\,...,\, Q_{q}\}$ are finite sets of irreducible polynomials
in Schwinger  parameters.  We assume furthermore  that all $P_i$ are homogeneous and positive or negative definite. This is the case 
for all Symanzik polynomials in the Euclidean momentum region and in the massless case, and also for the polynomials arising from their linear reduction 
in a large class of situations.
This simplifying assumption allows us to apply the particular change of variables constructed below. However, the general method is not restricted to this case.

A more precise description of the numerator
is given below. For our algorithms to be applicable we furthermore
have to assume that there is an ordering on the Schwinger  parameters
such that the set $\mathcal{P}$ is linearly reducible and unramified \cite{Bro5,Bro4}. In the following let  
$\alpha_N, \alpha_{N-1},...,\alpha_1$ be such a fixed ordering.

\subsection{From Schwinger  parameters to cubical variables\label{sub:From-Feynman-parameters}}

In the following, we transform a given integrand $f$ of the type given by $(\ref{hypIntegrand})$
to an integrand in cubical variables. According to our fixed ordering on the Schwinger  parameters, let $\alpha_N$ be the 
parameter to be integrated out in the present step. Linear
reducibility implies that the polynomials in $\mathcal{P}$ are of
degree at most 1 in $\alpha_{N},$ while there are no implications
for $\mathcal{Q}.$ We write $\mathcal{P}=\mathcal{P}_{N}\cup\mathcal{P}_{\backslash N}$
where $\mathcal{P}_{N}\subset\mathcal{P}$ is the subset of polynomials
linear in $\alpha_{N}$ and $\mathcal{P}_{\backslash N}\subset\mathcal{P}$
is the set of polynomials independent of $\alpha_{N}.$ Let us fix
the numbering on the $P_{i}$ such that $\mathcal{P}_{N}=\left\{ P_{1},\,...,\, P_{n}\right\} $
with $n\leq r.$ We also write the set of all polynomials $Q_{i}$
as $\mathcal{Q}=\mathcal{Q}_{N}\cup\mathcal{Q}_{\backslash N}$ where
the polynomials in $\mathcal{Q}_{N}$ depend on $\alpha_{N}$ and
the ones in $\mathcal{Q}_{\backslash N}$ do not.

Now let us be more specific  about the functions occurring in the numerator
of $(\ref{hypIntegrand})$. We write $L_{w}(\alpha_{N})$ for a
hyperlogarithm in $\alpha_{N},$ given by a word $w$
in differential 1-forms in the alphabet 
\begin{equation}
\Omega_{N}^{\textrm{Feynman}}=\left\{ \frac{d\alpha_{N}}{\alpha_{N}},\,\frac{d\alpha_{N}}{\alpha_{N}-\rho_{i}}\textrm{ where }\rho_{i}=
-\frac{ P_{i}|_{\alpha_{N}=0} }{\frac{\partial P_{i}}{\partial \alpha_{N}}}\textrm{ for }i=1,\,...,\, n\right\} .\label{eq: Omega Feyn}
\end{equation}
 Here $\rho_{i}$ is a rational function such that  $P_{i}$ vanishes for $\alpha_N=\rho_i$. 
Throughout this section, we shall assume the Feynman integral we are considering is linearly reducible and unramified. The condition for being unramified was defined in 
\cite{Bro5}, definition 16, and discussed in \cite{Bro4}, \S9.3. It implies in particular that if $\rho_i$ is a constant independent of all $\alpha_i$, then it must be equal to $0$ or $-1$.

We assume as an induction hypothesis that the functions in the numerator of the integrand are of a certain type. 
We will see in section \ref{sub:Back-to-Feynman} that this assumption will be satisfied 
after  integration and will be  the starting point for the next integration.  
The numerator of the integrand $f$ is assumed to be a linear
combination of hyperlogarithms in $\alpha_{N}:$ 
\begin{equation}
\textrm{numerator}(f)=\sum_{k}a_{k}b_{k}(\alpha_{N})L_{w_{k}}(\alpha_{N}),\label{eq:Feyn integrand hyperlog}
\end{equation}
where the $w_{k}$ are words in the alphabet $\Omega_{N}^{\textrm{Feynman}}$
and where we denote the $\alpha_{N}$-dependent and $\alpha_{N}$-independent
factor of the $k$-th coefficient by $b_{k}(\alpha_{N})$ and $a_{k}$
respectively. The $\alpha_{N}$-dependent factor $b_{k}(\alpha_{N})$
is a product of $Q_{i}\in\mathcal{Q}_{N}$ while the $\alpha_{N}$-independent
factor $a_{k}$ is allowed to be a product of $Q_{i}\in\mathcal{Q}_{\backslash N}$
and of hyperlogarithms which do not depend on $\alpha_{N}.$ As $\alpha_{N}$-independent
factors of the numerator remain unchanged in the integration procedure,
we restrict our attention to integrals of the type 
\begin{equation} \label{eq:LIntegral}
\int_{0}^{\infty}d\alpha_{N}f(\alpha_{1},\,...,\,\alpha_{N})=\int_{0}^{\infty}d\alpha_{N}\frac{\prod_{Q_{i}\in\mathcal{Q}_{N}}
Q_{i}^{\delta_{i}}L_{w}(\alpha_{N})}{\prod_{P_{i}\in\mathcal{P}}P_{i}^{\beta_{i}}}.
\end{equation}

Let us now express the integral of $(\ref{eq:LIntegral})$ as
an integral over cubical coordinates such that the algorithms of section
\ref{sec:Computing-on-the} apply. Let $\mathbb{R}_{+}^{N}$ be the
subspace of $\mathbb{R}^{N}$ where all Schwinger  parameters are greater than
or equal to zero and let $\mathbb{R}_{\textrm{cube}}^{n}$ be the unit cube
in $n$ cubical variables, i.e. 
\begin{eqnarray*}
\mathbb{R}_{+}^{N} & = & \left\{ (\alpha_{1},...,\,\alpha_{N})\in\mathbb{R}^{N}|0\leq\alpha_{i},\, i=1,\,...,\, N\right\} ,\\
\mathbb{R}_{\textrm{cube}}^{n} & = & \left\{ (x_{1},...,\, x_{n})\in\mathbb{R}_{n}|0\leq x_{i}\leq1,\, i=1,\,...,\, n\right\} .
\end{eqnarray*}

 Consider the $\alpha_{N}$-dependent polynomials $\mathcal{P}_{N}=\{P_{1},\,...,\, P_{n}\}$
and the corresponding $\rho_{i}=-\frac{P_{i}|_{\alpha_{N}=0}}{ \frac{\partial P_{i}}{\partial\alpha_{N}}  }$ for $i=1,\,...,\, n.$
We introduce an ordering on the set $\mathcal{P}_{N}$ as follows. A sufficiently small open region of the form  
$0\leq \alpha_{N-1} \ll \alpha_{N-2} \cdots \ll \alpha_1 \ll \epsilon$ (where $x\ll y $ denotes $x<y^M$ for some large $M$) does not intersect the hypersurfaces $\rho_i-\rho_j=0$. Therefore  number the polynomials in $\mathcal{P}_{N}=\{P_{1},\,...,\, P_{n}\}$ such that 
everywhere in  this region we have
\begin{equation}
0 > \rho_{n}>\rho_{1}>\rho_{2}>...>\rho_{n-2}>\rho_{n-1}.\label{eq:ordered zeroes}
\end{equation}

For the given, ordered set $(P_{1},\,...,\, P_{n})$, consider the 
rational map between affine spaces
\begin{equation*}
\phi:\,\mathbb{A}^{N}\rightarrow\mathbb{A}^{n},\ 
\end{equation*}
(equivalently a  homomorphism
$\phi^*: \mathbb{Q}(x_1,\ldots, x_n) \rightarrow \mathbb{Q}(\alpha_1,\ldots, \alpha_N)$)
 given by 
 \begin{eqnarray}
\phi^*(x_{n}) & = & \frac{\alpha_{N}}{\alpha_{N}-\rho_{n}}\nonumber, \\
\phi^*(x_{n-1}) & = & 1-\frac{\rho_{n}}{\rho_{n-1}},\nonumber \\
\phi^*(x_{k}) & = & \frac{1-\frac{\rho_{n}}{\rho_{k}}}{1-\frac{\rho_{n}}{\rho_{k+1}}}\textrm{ for }1\leq k\leq n-2.\label{eq: change variables}
\end{eqnarray}
These variables $x_i$ will be our cubical coordinates and we construct the set of 1-forms $\bar{\Omega}^F_n$ as above. 
Note that the restriction of $\phi$ to the first $N-1$ coordinates defines a rational map $\phi: \mathbb{A}^{N-1} \rightarrow \mathbb{A}^{n-1}$, 
since $\rho_1,\ldots, \rho_{n}$ do not depend on $\alpha_N$.
For fixed $\alpha_1,\ldots, \alpha_{N-1}$, the curve $\Pro^1$ with coordinate $\alpha_N$ and punctures at $\{0,\rho_1,\ldots, \rho_{n},\infty \}$ (i.e., 
the fiber of $\mathbb{A}^N \rightarrow \mathbb{A}^{N-1}$), 
 is isomorphic, via  $\ref{eq: change variables}$, 
to the curve with coordinate $x_n$ and punctures at $\{0,(x_1\ldots x_{n-1})^{-1}, (x_2\ldots x_{n-1})^{-1} , \ldots , x_{n-1}^{-1},\infty, 1\}$, in that order.
 Via such a (family of) isomorphisms, we can  explicitly express 
all 1-forms in $\Omega_{N}^{\textrm{Feynman}}$ as $\mathbb{Q}$-linear combinations of  1-forms in $\bar{\Omega}^F_n$ in cubical coordinates. We obtain 
\begin{eqnarray} \label{dalphas}
\frac{d\alpha_N}{\alpha_N} & = &  \frac{dx_n}{x_n} - \frac{dx_n}{x_n-1} ,\\
\frac{d\alpha_N}{\alpha_N-\rho_n} & = & -\frac{dx_n}{x_n-1}  \ ,   \nonumber \\ 
\frac{d\alpha_N}{\alpha_N-\rho_i} & = &  \frac{x_i ... x_{n-1}dx_n}{x_i... x_n-1} - \frac{dx_n}{x_n-1} ,  \nonumber  
\end{eqnarray}
since the $\rho_i$ are constant, for $i=1,...,n-1$. As a consequence, we can express each hyperlogarithm $L_{w}(\alpha_{N})$
as a $\mathbb{Q}$-linear combination of hyperlogarithms in cubical variables $\xi\in V\left(\bar{\Omega}^F_{n}\right).$ 

For simplicity, we  make the following assumption (which is slightly stronger than assuming that  the linear reduction of the Feynman integral is unramified):
\begin{equation}
 \label{eq:xlimits}
\lim_{\alpha_{1}\rightarrow0}...\lim_{\alpha_{N}\rightarrow0}x_{k}(\alpha_{1},\,...,\,\alpha_{N})\in\{0,\,1\},\, k=1,\,...,\, n,
\end{equation}
where these limits are approached from inside the cube $\mathbb{R}_{\textrm{cube}}^{n}.$
 The domain of the $\alpha_{N}$-integration is mapped to the domain
$0\leq x_{n}\leq1.$ The Jacobian is  $J=-\frac{\rho_{n}}{(x_{n}-1)^{2}}.$

Up to rational functions which do not depend on $x_{n},$ we can now express integrals of the type
of $(\ref{eq:LIntegral})$ as integrals of the type    
\begin{equation} \label{eq:IntCubical}
\int_{0}^{1}dx_{n}\frac{\prod q_{i}^{\gamma_{i}}}{\prod f_{i}^{\delta_i}} \xi
\end{equation}
where $\gamma_{i},\delta_{i}\in\mathbb{N},$ and where each $q_{i}$ is a polynomial in Schwinger  parameters without
$\alpha_{N}$ or in cubical variables, and the integrand involves functions $f_{i}\in\{x_{n},\, x_{n}-1,\, x_{n-1}x_{n}-1,\,...,\, x_{1}\cdot\cdot\cdot x_{n}-1\}$
and hyperlogarithms $\xi\in V\left(\bar{\Omega}^F_{n}\right).$ Before we can apply our algorithm of subsection \ref{sub:Primitives} for 
the computation of primitives, we use a standard procedure of applying finitely many successive partial fraction decompositions
 and partial integrations until all powers $\delta_i$ are equal to 1. 

As a last step of preparation, we apply the symbol map $\Psi$ of subsection
\ref{sub:The-symbol-map} to $\xi$. We obtain an integral as in $(\ref{eq:IntCubical})$ where now $\xi\in V\left(\Omega_{n}\right).$
Now we compute the definite integral $(\ref{eq:IntCubical})$ by use
of the algorithms of subsections \ref{sub:Primitives} and \ref{sub:Limits}.
Up to rational prefactors, we obtain a $\mathcal{Z}$-linear combination
of functions in $V\left(\Omega_{n-1}\right).$

\subsection{Back to Schwinger  parameters\label{sub:Back-to-Feynman} }

Note that after the integration, we have a function in terms of both types of variables, the Schwinger  parameters and the cubical coordinates. 
In order to proceed with the integration over a next Schwinger  parameter and apply the same steps again,
we firstly have to express the integrand only in terms of  Schwinger  parameters again.
Let $I$ be the result of the $\alpha_{N}$-integration, expressed
as a linear combination
\[
I=\sum a_{i}\xi_{i}
\]
of multiple polylogarithms $\xi_{i}\in V\left(\Omega_{n-1}\right)$.
The coefficients $a_{i}$ are trivially expressed by Schwinger  parameters
by application of $\phi^*.$ However, expressing the multiple polylogarithms
$\xi_{i}$ in terms of Schwinger  parameters is more subtle, as we have
to respect the limiting conditions of iterated integrals in both sets
of variables.

For any function $f$ of variables $y_{1},\,...,\, y_{n}$ and numbers
$c_{1},\,...,\, c_{n}$ let us introduce the notation
\[
\lim_{(y_{1},\,...,\, y_{n})\rightarrow(c_{1},\,...,\, c_{n})}f=\mathrm{Reg}_{y_{n}\rightarrow c_{n}}...\mathrm{Reg}_{y_{1}\rightarrow c_{1}}f.
\]
where in the right-hand side, $\mathrm{Reg}$ denotes the regularised limits with respect to unit tangent vectors in either cubical coordinates $x_i$ (or $1-x_i$), or 
Schwinger parameters $\alpha_i$.
In the following let us write $0_{n}$ for the vector $(0,\,...,\,0)$
with $n$ components. We consider the vector $x_{p}=(x{}_{p(1)},\,...,\, x_{p(n-1)})$
of the remaining cubical coordinates, where the ordering is given by a permutation $p$ on the set $\{1,\,...,\, n-1\}$.
We furthermore consider the vector of remaining Schwinger  parameters $\alpha=(\alpha{}_{N-1},\,...,\,\alpha_{1})$ in the 
ordering in which we integrate over them, as fixed above. 

Consider a multiple polylogarithm $\xi\in V\left(\Omega_{n-1}\right).$
By definition, it satisfies 
\begin{equation}
\lim_{x_{\sigma}\rightarrow0_{n-1}}\xi= \epsilon ( \xi) \label{eq:limit condition cubical}
\end{equation}
for every permutation $\sigma$ on $\{1,\,...,\, n-1\}$, where $\epsilon$ is the augmentation map (projection onto components of length $0$). We want to express
each $\xi$ as iterated integrals $\eta$ in Schwinger  parameters, for
which we impose the condition 
\begin{equation}
\lim_{\alpha\rightarrow0_{N-1}}\eta=\epsilon ( \eta) . \label{eq:limit condition Feynman}
\end{equation}

Condition $\ref{eq:limit condition Feynman}$ corresponds to 
a vanishing condition for the iterated integral $\xi \in V\left(\Omega_{n-1}\right)  $ at a tangential base  point on   $\Mod_{0,n+2}$  (strictly speaking, 
on a related space $\Mod_{0,n+2}^{\dag}$ (\cite{Bro4}, \S8.2) which  can be read off from the linear reduction algorithm and  involves removing from $\mathbb{A}^{n-1}$ only those hypersurfaces $x_i=0$, $x_ix_{i+1}\ldots x_j =1$ which correspond to singularities actually occurring in the integrand),
which is on the boundary of the connected component of $\Mod_{0,n+2}(\mathbb{R})$ defined by the unit hypercube $0\leq x_1,\ldots, x_{n-1} \leq 1$.
One can verify that such  a point can always be  represented by a permutation 
$p$ on $\{1,\,...,\, n-1\}$ (non-uniquely) and a vector $c=(c_{1},\,...,\, c_{n-1})$ (uniquely)
with all $c_{i}\in\{0,\,1\}$ such that  for any rational function
$g$ in the $x_i$ which is regular  on $\Mod^{\dag}_{0,n+2}$, we have 
\begin{equation}
\lim_{x_{p}\rightarrow c}g=\lim_{\alpha\rightarrow0_{N-1}}\phi^\star g,\label{eq:limits}
\end{equation}
where on the left-hand side $c$ is approached inside $\mathbb{R}_{\textrm{cube}}^{n-1}$
and on the right-hand side $(0,\,...,\,0)$ is approached inside $\mathbb{R}_{+}^{N-1}.$ Such a point $c$ and permutation $p$ determine the procedure 
to express $\xi$ in terms of Schwinger  parameters.
The components of $c$ are computed by
\[
c_{i}=\lim_{\alpha \rightarrow0_{N-1}}x_{i},
\]
where $i\in\{1,\,...,\, n-1\}$, and lies in $\{0,1\}$, by assumption $(\ref{eq:xlimits})$.
 In the  case when $\Mod_{0,n+2}^{\dag}= \Mod_{0,n+2}$ (i.e., all  possible singularities which can  occur actually do occur), a permutation $p$ satisfying 
$(\ref{eq:limits})$ can easily be computed with the help of dihedral coordinates $u_{ij}$, which are related to the cubical coordinates as discussed in \cite{Bro2}. 
A permutation $p$ satisfies $(\ref{eq:limits})$ for any regular function $g$ on $\Mod_{0,n+2}$ (expressed as a rational function of cubical coordinates) if it satisfies 
\[
\lim_{x_{p}\rightarrow c}u_{ij}=\lim_{\alpha \rightarrow0_{N-1}}\phi^\star u_{ij}
\]
for all dihedral coordinates $u_{ij}$.  This condition determines $p$  in this case.

\begin{example}
Suppose  $\Mod_{0,5}^{\dag}=\Mod_{0,5}$. Let $x_1,x_2$ be cubical coordinates. Suppose that    $x_1=1-\alpha_2, x_2=1-{\alpha_2 \over \alpha_1}$.
Then the five dihedral coordinates $(x_1,x_2, 1-x_1x_2, {1-x_1 \over 1-x_1x_2}, {1-x_2 \over 1-x_1x_2})$ in the limits $\alpha_2\rightarrow 0$ then $\alpha_1 \rightarrow 0$ tend to $(1,1,0,0,1)$ respectively. This corresponds to taking first the limit as $x_1\rightarrow 1$ and then $x_2 \rightarrow 1$.

  On the other hand, suppose that $x_1 = 1-\alpha_1, x_2 = 1-\alpha_1$. Then the limit of the five dihedral coordinates above as $\alpha_1 \rightarrow 0$ are
  $(1,1,0,{1\over 2}, {1\over 2})$, which could potentially produce a $\log 2$ in the iterated integrals (ramification at prime $2$). In such a case, the condition of being unramified  will ensure that $1-x_1x_2=\alpha_1(2-\alpha_1)$ does not occur as a singularity of the integrand. Thus $\Mod_{0,5}^{\dag} = \mathbb{A}^2 \backslash \{x_1,x_2=0,1\} = \Mod_{0,4} \times \Mod_{0,4}$ strictly contains $\Mod_{0,5}$.
 The limit as $\alpha_1\rightarrow 0$ can be obtained as the limit as  $x_1\rightarrow 1, x_2\rightarrow 1$ in either order.
\end{example}

Now let $x_{p}$ and $c$ be vectors satisfying $(\ref{eq:limits})$.
We define $\eta$ by the following equation, where $\xi\in V\left(\Omega_{n-1}\right)$
is the result of the integration of $(\ref{eq:IntCubical})$,
\begin{equation}
\eta=m\left(\phi^{\star}\otimes\phi^{\star}\lim_{x_{p}\rightarrow c}\right)\Delta\xi \label{eq:limit procedure}
\end{equation}
and  $m$ is multiplication. Note that this is an application of $(\ref{pathconcat})$.
Then $\eta $  is the desired expression in terms of Schwinger 
parameters.

As a last step, we express each iterated integral in terms of hyperlogarithms,
such that we arrive at the starting point for the next integration over the variable $\alpha_{N-1}$.
As a consequence of linear reducibility, all iterated integrals
are now given by differential forms of the type $\omega=dP/P$ where $P$
are polynomials in the Schwinger  parameters which are of degree $\leq1$
in the variable $\alpha_{N-1}$. In analogy to the construction of the unshuffle map we define the auxiliary restriction operations
\[
\pi_{\alpha_{i}}\omega=\omega|_{d\alpha_{i}=0,\,\alpha_{i}=0}
\]
and 
\[
r_{\alpha_{i}}\omega=\omega|_{d\alpha_{j}=0\textrm{ for all }j\neq i.}
\]
By 
\begin{equation}
\eta'=m\left(r_{\alpha_{N-1}}\otimes \pi_{\alpha_{N-1}}\right)\Delta \eta \label{eq:write hyperlog}
\end{equation}
we finally arrive at a linear combination of hyperlogarithms $L_{w}(\alpha_{N-1})$
whose coefficients are products of rational functions in Schwinger  parameters,
multiple zeta values and iterated integrals independent of $\alpha_{N-1}.$
Iterating the computation of $(\ref{eq:write hyperlog})$ for the
remaining Schwinger  parameters we can express all iterated integrals
as hyperlogarithms. With this expression we can repeat the above steps
to integrate out $\alpha_{N-1}$, and so on.

\subsection{Summary of the integration algorithm}

Let us summarize the above steps for integrating over one Schwinger 
parameter $\alpha_{N}.$ We start from a finite integral $I=\int_{0}^{\infty}d\alpha_{N}f$
whose integrand $f$, as in $(\ref{eq:Feyn integrand hyperlog})$, is
a linear combination of hyperlogarithms $L_{w}(\alpha_{N})$ as functions
of $\alpha_{N}$, and whose coefficients are products of rational functions
$b(\alpha_{N})$ of the Schwinger  parameters including $\alpha_{N},$
and further functions (possibly hyperlogarithms) not depending on
$\alpha_{N}.$ As above, we write $\mathcal{P}_{N}$ for the set of
$n$ polynomials depending linearly on $\alpha_{N},$ which are in
the denominators of $b(\alpha_{N})$ and define the differential forms
of $L_{w}(\alpha_{N})$ by $(\ref{eq: Omega Feyn})$. The set $\mathcal{P}_{N}$
is linearly reducible with respect to an ordered set $(\alpha_N,...,\alpha_1)$ of
all Schwinger  parameters, and unramified.

The main steps of the algorithm are combined as follows: 
\begin{itemize}
\item Define the $n$ cubical variables $x_{1},\,...,\, x_{n}$, and express the integrand
 $f$ via the map     $\ref{dalphas}$   as a linear combination
of hyperlogarithms in $V\left(\bar{\Omega}^F_{n}\right)$. The integration
over $\alpha_{N}$ is mapped via \ref{eq: change variables}  to the integration over $x_{n}$ from
0 to 1.
\item Apply the symbol map $\Psi$ of subsection \ref{sub:The-symbol-map}
to lift each function in $V\left(\bar{\Omega}^F_{n}\right)$ to multiple
polylogarithms in $V\left(\Omega_{n}\right).$
\item Use iterated partial integration and partial fraction decomposition
to bring the integrand into the appropriate form. Then use the map $\star$
of subsection \ref{sub:Primitives} to compute the primitive of $f.$
\item Take the limits of the primitive at $x_{n}=0$ and $x_{n}=1$ to obtain
the definite integral from 0 to 1, using the algorithm of subsection
\ref{sub:Limits}. The result is a linear combination of multiple
polylogarithms in $V\left(\Omega_{n-1}\right)$ with coefficients
possibly involving multiple zeta values.
\item Apply the change of variables to obtain an expression only in Schwinger  parameters again. For iterated integrals, apply $(\ref{eq:limit procedure})$
such that the regularised limit at $\alpha\rightarrow0_{N-1}$ is preserved.
\item Write the result as a combination of hyperlogarithms in the next integration
variable by $(\ref{eq:write hyperlog})$.
\end{itemize}
Examples for the application of this algorithm by use of our computer
program are given below.

\section{Applications}\label{sec:Applications}

\subsection{Cellular integrals}
A particular instance of period integrals on moduli spaces are given by the cellular integrals defined in \cite{AperyVar} in relation to irrationality proofs.
The basic construction is to consider a permutation $\sigma $ of $\{1,\ldots, n\}$ and define a rational function and differential form
 $$
 \widetilde{f}_{\sigma}   =  \prod_{i} {z_i -z_{i+1} \over z_{\sigma(i)} - z_{\sigma(i+1)} }
 \qquad \hbox{ and } \qquad  \widetilde{\omega}_{\sigma}  =  \prod_{i} {dz_i\over z_{\sigma(i)} - z_{\sigma(i+1)} } \nonumber\ ,
  $$
 on the configuration space  $C^n (\Pro^1)$ of $n$ distinct points $z_1,\ldots, z_n$ in $\Pro^1$,
 where the product is over all indices $i$ modulo $n$.
 Now $\mathrm{PGL}_2$ acts diagonally on $C^n (\Pro^1)$, and the quotient is 
 $$\mathcal{M}_{0,n} \cong C^n( \Pro^1) / \mathrm{PGL}_2\ .$$
 The rational function $\widetilde{f}_\sigma$ and the form $\widetilde{\omega}_{\sigma}$ are $\mathrm{PGL}_2$-invariant, and therefore descend in the standard way to 
 a  rational function and form $f_{\sigma}, \omega_{\sigma}$ on $\mathcal{M}_{0,n}$. Because $\mathrm{PGL}_2$ is triply-transitive, we can  put $z_1=0, z_{n-1}=1, z_n = \infty$, and 
 replace $z_{i+1}$ by $x_ix_{i+1}\ldots x_{n-3}$ for $i=1,\ldots, n-3$, where $x_1,\ldots, x_{n-3}$ are cubical coordinates on $\Mod_{0,n}$. 
 
 Therefore we can formally  write
  $$ f_{\sigma}   =  \prod_{i} {z_i -z_{i+1} \over z_{\sigma(i)} - z_{\sigma(i+1)} }    
   \qquad \hbox{ and } \qquad  
     \omega_{\sigma}  =     { dx_1\ldots dx_{n-3} \over \prod_i  z_{\sigma(i)} - z_{\sigma(i+1)} } \nonumber\ , $$
 where the product is over all indices $i$ modulo $n$, and all factors involving $z_n=\infty$ are simply omitted. For all $N\geq 0$, consider the family of basic cellular  integrals
 \begin{equation}
 I_N^{\sigma} = \int_{[0,1]^{n-3}} f_{\sigma}^N \omega_{\sigma}
 \end{equation} 
 where the domain of integration is the unit hypercube in the cubical coordinates $x_i$.
 Conditions for the convergence of the integral are discussed in \cite{AperyVar}.
When it converges, this integral is a rational linear combination of multiple zeta values of weights $\leq n-3$ and can be computed with our program.
In the case $n=5,6$ and $\sigma(1,2,3,4,5) = (1,3,5,2,4)$, and $\sigma(1,2,3,4,5,6)=(1,3,6,4,2,5) $ it gives back exactly the linear forms involved in Ap\'ery's proofs of the irrationality of $\zeta(2)$ and $\zeta(3)$.
A systematic study of examples for higher $n$ (described in \cite{AperyVar}) was undertaken using the algorithms described in this paper.

\subsection{Expansion of generalized hypergeometric functions}

Many Feynman integrals can be expressed in terms of generalized hypergeometric
functions
\[
_{p}F_{q}(a_{1},\,...,\, a_{p};\, b_{1},\,...,\, b_{q};\, z)=\sum_{k=0}^{\infty}\frac{\prod_{j=1}^{p}\left(a_{j}\right)_{k}z^{k}}{\prod_{j=1}^{q}\left(b_{j}\right)_{k}k!},
\]
 converging everywhere in the $z$-plane if $q\geq p,$ and in the
case $q=p-1$ for $|z|<1$ or at $|z|=1$ if the real part of $\sum_{j=1}^{p-1}b_{j}-\sum_{j=1}^{p}a_{j}$
is positive. Here we used the Pochhammer symbol
 \begin{equation*}
(a)_n=\frac{\Gamma (a+n)}{\Gamma (a)}.
 \end{equation*}
Multi-variable generalizations, such as Appell and
Lauricella functions, play a role in Feynman integral computations as well.
If the Feynman integral is considered in $D=4-2\epsilon$ dimensions, the parameters take the form 
\begin{equation} \label{eq:HyperParameters}
a_{i}=A_{i}+\epsilon\alpha_{i},\, b_{i}=B_{i}+\epsilon\beta_{i}\;\textrm{where }\alpha_{i},\,\beta_{i}\in\mathbb{R}.
\end{equation}
In the case of massless integrals, the numbers $A_{i},\, B_{i}$ are
integers while in the case of non-vanishing masses, some of them are
half-integers. 

In order to arrive at a result for the Feynman integral where pole-terms
in $\epsilon$ can be separated, one has to expand these functions
near $\epsilon=0.$ Several computer programs are available for this task.
The programs of \cite{MocUwe, Wei} use algorithms for the expansion of very
general types of nested sums \cite{MocUweWei} while the program \cite{HubMai2}
writes an Ansatz in harmonic polylogarithms and determines the coefficients
from differential equations. A method using systems of differential
equations was presented in \cite{Kal, KalWarYos1, KalWarYos2}. We also 
refer to \cite{Kal2, Grey, AblX} for recent progress in the field.

Alternatively, we can start from an integral representation of the
function, expand the integrand and compute the resulting integrals
explicitly. This approach was applied in \cite{HubMai1} for the expansion
of $_{2}F_{1}.$ The algorithms presented above are very well suited
for this method and can be used to extend it to more general functions. 

As a first example we still consider $_{2}F_{1}.$ We have the integral
representation
\[
_{2}F_{1}(a_{1},\, a_{2};\, b;\, z_{1})=\frac{\Gamma(b)}{\Gamma(a_{2})\Gamma(b-a_{2})}\int_{0}^{1}z_{2}^{a_{2}-1}(1-z_{2})^{b-a_{2}-1}(1-z_{1}z_{2})^{-a_{1}}dz_{2}
\]
for $\textrm{Re}(b)>\textrm{Re}(a_{2})>0$ and $z_1\notin [1,\infty).$ The parameters $a_i$ and $b$ may depend on $\epsilon$ as in $(\ref{eq:HyperParameters})$.
If we exclude the case of half-integers mentioned above, the expansion at $\epsilon=0$ leads to integrands whose denominators are products of $z_2, (1-z_2), (1-z_1 z_2)$ and whose numerators may involve powers of 
logarithms of these functions. We can view the variables $z_1, z_2$ as cubical coordinates and apply the algorithms of section \ref{sec:Computing-on-the} to integrate over
$z_{2}$ analytically. 

Example: 
\begin{eqnarray*}
_{2}F_{1}(1,\,1+\epsilon;\,3+\epsilon;\, z_{1}) & = & \frac{\Gamma(3+\epsilon)}{\Gamma(1+\epsilon)}\int_{0}^{1}\frac{z_{2}^{\epsilon}(1-z_{2})}{1-z_{1}z_{2}}dz_{2}\\
 & = & \int_{0}^{1}\frac{2(z_{2}-1)}{z_{1}z_{2}-1}dz_{2}+\epsilon\int_{0}^{1}\frac{\left(2\ln(z_{2})+3\right)(z_{2}-1)}{z_{1}z_{2}-1}\\
 &  & +\epsilon^{2}\int_{0}^{1}\frac{\left(\ln(z_{2})^{2}+3\ln(z_{2})+1\right)(z_{2}-1)}{z_{1}z_{2}-1}dz_{2}+\mathcal{O}\left(\epsilon^{3}\right)\\
 & = & \frac{2}{z_{1}^{2}}\left(z_{1}+(1-z_{1})\ln(1-z_{1})\right)+\frac{\epsilon}{z_{1}^{2}}\left(z_{1}+3(1-z_{1})\ln(1-z_{1})\right.\\
 &  & \left.+2(1-z_{1})\textrm{Li}_{2}(z_{1})\right)+\frac{\epsilon^{2}}{z_{1}^{2}}\left((1-z_{1})\ln(1-z_{1})\right.\\
 &  & \left.+3(1-z_{1})\textrm{Li}_{2}(z_{1})-2(1-z_{1})\textrm{Li}_{3}(z_{1})\right)+\mathcal{O}\left(\epsilon^{3}\right)
\end{eqnarray*}

We extend the approach to generalized hypergeometric functions, starting
from the integral representation
\[
_{p}F_{q}(a_{1},\,...,\, a_{p};\, b_{1},\,...,\, b_{q};\, z)
\]
\[
=\frac{\Gamma(b_{q})}{\Gamma(a_{p})\Gamma(b_{q}-a_{p})}\int_{0}^{1}t^{a_{p}-1}(1-t)^{b_{q}-a_{p}-1}\,_{p-1}F_{q-1}(a_{1},\,...,\, a_{p-1};\, b_{1},\,...,\, b_{q-1};\, zt)dt
\]
in the region where it converges. Here again the expansion of the integrand leads to integrals over cubical coordinates which can be computed by the algorithms of section \ref{sec:Computing-on-the}.

Example: 
\begin{eqnarray*}
_{3}F_{2}(2,\,1+\epsilon,\,1+\epsilon;\,3+\epsilon,\,2+\epsilon;\, z_{1}) & = & \frac{\Gamma(3+\epsilon)\Gamma(2+\epsilon)}{\Gamma(1+\epsilon)^{2}}\int_{0}^{1}\int_{0}^{1}\frac{z_{2}z_{3}^{\epsilon}(1-z_{2})^{\epsilon}}{(1-z_{1}z_{2}z_{3})^{1+\epsilon}}dz_{2}dz_{3}\\
 & = & \frac{2}{z_{1}^{2}}\left((1-z_{1})\ln(1-z_{1})+z_{1}\right)+\frac{\epsilon}{z_{1}^{2}}\left(7(1-z_{1})\ln(1-z_{1})\right.\\
 &  & \left.+5z_{1}+(2-4z_{1})\textrm{Li}_{2}(z_{1})\right)+\frac{\epsilon^{2}}{z_{1}^{2}}\left(9(1-z_{1})\ln(1-z_{1})\right.\\
 &  & \left.+(7-12z_{1})\textrm{Li}_{2}(z_{1})+(6z_{1}-2)\textrm{Li}_{3}(z_{1})+4z_{1}\right)+\mathcal{O}\left(\epsilon^{3}\right)
\end{eqnarray*}

While for these functions the integral representations are readily
given in cubical coordinates, an extension to further cases may require
a change of variables. For example the first Appell function 
\[
F_{1}(a;\, b_{1},\, b_{2};\, c;\, x,\, y)=\sum_{m\geq0}\sum_{n\geq0}\frac{\left(a\right)_{m+n}\left(b_{1}\right)_{m}\left(b_{2}\right)_{n}}{m!n!\left(c\right)_{m+n}}x^{m}y^{n}\;\textrm{where }|x|,\,|y|<1
\]
with the integral representation \cite{Pic} 
\[
F_{1}(a;\, b_{1},\, b_{2};\, c;\, x,\, y)=\frac{\Gamma\left(c\right)}{\Gamma(a)\Gamma(c-a)}\int_{0}^{1}t^{a-1}(1-t)^{c-a-1}(1-tx)^{-b_{1}}(1-ty)^{-b_{2}}dt
\]

for $\textrm{Re}(c)>\textrm{Re}(a)>0$ can be expressed in the appropriate
form after introducing cubical coordinates $z_{3}=t,$ $z_{2}=y,$
$z_{1}=x/y.$

Example: 
\begin{eqnarray*}
F_{1}(1;\, 1,\, 1;\, 2+\epsilon;\, x,\, y) & = & \frac{\Gamma(2+\epsilon)}{\Gamma(1+\epsilon)}\int_{0}^{1}\frac{(1-z_{3})^{\epsilon}}{(1-z_{1}z_{2}z_{3})(1-z_{2}z_{3})}dz_{3}\\
 & = & \frac{1}{x-y}\left(\ln(1-y)-\ln(1-x)\right)\\
 &  & +\frac{\epsilon}{x-y}\left(\ln(1-y)-\ln(1-x)+\frac{1}{2}\ln(1-y)^{2}-\frac{1}{2}\ln(1-x)^{2}\right.\\
 &  & -\left.\textrm{Li}_{2}(x)+\textrm{Li}_{2}(y)\right)+\mathcal{O}\left(\epsilon^{2}\right)
\end{eqnarray*}

We checked the examples with $_{2}F_{1}$ and $_{3}F_{2}$ analytically with the program of \cite{HubMai2} and the example with $F_{1}$
numerically with the built-in first Appell function in Mathematica. 

\subsection{Feynman integrals}

As a third application we turn to the computation of Feynman integrals
by direct integration over their Schwinger parameters. As a first example we
consider the period integral (in the sense of \cite{BEK}) of the four-loop vacuum-type graph of
figure 5.1 a). The integral is finite in $D=4$ dimensions and is given
in terms of Schwinger parameters as
\[
I_{1}=\int_{\alpha_{i}\geq0}\prod_{i=1}^{8}d\alpha_{i}\delta(1-\alpha_{8})\frac{1}{\mathcal{U}^{2}}.
\]
The first Symanzik polynomial $\mathcal{U}$ is linearly reducible
in this case. We use our implementation of the algorithms of sections
\ref{sec:Computing-on-the} and \ref{sec:Feynman-type-integrals} to integrate over $\alpha_{1},\,...,\, \alpha_{7}$ in an appropriate
ordering and to compute the limit at $\alpha_{8}=1$ in the last step.

 The computation time per integration grows at first 
due to the increasing weight and complexity of the functions involved,
but decreases in the end as  fewer variables remain. Here we compute
with multiple polylogarithms of weight 2, 3, 4 and 5 in the fourth,
fifth, sixth and seventh integration respectively. We obtain the result
$I_{1}=20\zeta(5)$ which is well-known \cite{CheTka}. Period integrals
of this type appear as coefficients of two-point integrals corresponding
to graphs obtained from breaking open one edge in the vacuum-graph
(see \cite{Bro5, CheTka}).

As an example for a Feynman integral with non-trivial dependence on
masses and external momenta, we consider the hexagon-shaped one-loop
graph of figure 5.1 b) with incoming external momenta $p_{1},\,...,\, p_{6}.$ Introducing
one particle mass with $m^{2}<0$ we impose the on-shell condition
$p_{1}^{2}=m^{2},\, p_{i}^{2}=0,\, i=2,\,...,\,6.$ In $D=6$ dimensions
the Feynman integral reads 
\[
I_{2}=\int_{\alpha_{i}\geq0}\prod_{i=1}^{6}d\alpha_{i}\delta(1-\alpha_{6})\frac{2}{\mathcal{F}^{3}}
\]
with the second Symanzik polynomial
\[
\mathcal{F}=\sum_{1\leq i<j\leq 6}\alpha_{i}\alpha_{j}\left(-s_{ij}^{2}\right)
\]
and kinematical invariants 
\[
s_{ij}=\left(\sum_{k=i+1}^{j}p_{k}\right)^2.
\]
This integral is computed in \cite{Del3}. In a first step in this computation, the integral is expressed in terms of the cross-ratios 
\[
u_{1}=\frac{s_{26}^{2}s_{35}^{2}}{s_{25}^{2}s_{36}^{2}},\, u_{2}=\frac{s_{13}^{2}s_{46}^{2}}{s_{36}^{2}s_{14}^{2}},\, u_{3}=\frac{s_{15}^{2}s_{24}^{2}}{s_{14}^{2}s_{25}^{2}},\, u_{4}=\frac{s_{12}^{2}s_{36}^{2}}{s_{13}^{2}s_{26}^{2}}
\]
as 
\[
I_{2}=\frac{1}{s_{14}^{2}s_{25}^{2}s_{36}^{2}}\int_{\alpha_{i}\geq0}\prod_{i=1}^{3}d\alpha_{i}\frac{1}{(u_{2}+\alpha_{1}+\alpha_{2})(u_{3}\alpha_{1}+u_{1}\alpha_{3}+\alpha_{2})(u_{4}\alpha_{1}\alpha_{2}+\alpha_{2}+\alpha_{1}\alpha_{3}+\alpha_{3})}.
\]
We choose the parametrization 
\[
u_{1}=\frac{1}{1+y},\, u_{2}=\frac{1+v}{1+v-u},\, u_{3}=\frac{(1-u)(-y-x)}{(1+y)(-1+u-v)},\, u_{4}=\frac{1+v-x}{1+v}
\]
which differs from the one in \cite{Del3}. This parametrisation is not pulled from thin air: it is constructed recursively
out of the polynomials occurring in the linear reduction algorithm, applied to the integrand.
 With this choice each $u_{i}$
tends to either 0 or 1 at the tangential base-point which we choose by the ordering
$(\alpha_{2},\, \alpha_{3},\, \alpha_{1},\, u,\, v,\, x,\, y)$ and furthermore the
polynomials in the denominator of the re-written integrand of $I_{2}$
are linearly reducible for the ordering $(\alpha_{2},\, \alpha_{3},\, \alpha_{1})$.
Therefore we can apply our implementation to integrate over the $\alpha_{i}$
in this order and we obtain a result for positive $u,\, v,\, x,\, y$.
We checked the result analytically with the program of \cite{Pan3}. 
\begin{figure}
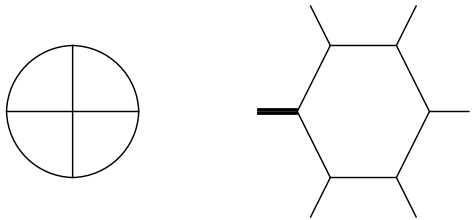
 \label{fig:Graphs}
\begin{centering} 
\begin{feynartspicture}(220, 120)(2, 1) 
\FADiagram{}
\FAProp(5.,10.)(10.,15.)(-0.4,){Straight}{0}  \FAProp(10.,15.)(15.,10.)(-0.4,){Straight}{0}  \FAProp(15.,10.)(10.,5.)(-0.4,){Straight}{0}  \FAProp(10.,5.)(5.,10.)(-0.4,){Straight}{0}  \FAProp(5.,10.)(10,10.)(0.,){Straight}{0}  \FAProp(10.,15.)(10.,10.)(0.,){Straight}{0}  \FAProp(15.,10.)(10.,10.)(0.,){Straight}{0}  \FAProp(10.,5.)(10.,10.)(0.,){Straight}{0}  

\FADiagram{}
\FAProp(7.5,15.)(12.5,15.)(0.,){Straight}{0}  \FAProp(12.5,15.)(15.,10.)(0.,){Straight}{0}  \FAProp(15.,10.)(12.5,5.)(0.,){Straight}{0}  \FAProp(12.5,5.)(7.5,5.)(0.,){Straight}{0}  \FAProp(7.5,5.)(5.,10.)(0.,){Straight}{0}  \FAProp(5.,10.)(7.5,15.)(0.,){Straight}{0}  \FAProp(15.,10.)(18.,10.)(0.,){Straight}{0}  \FAProp(12.5,15.)(14.,18.)(0.,){Straight}{0}  \FAProp(7.5,15.)(6.,18.)(0.,){Straight}{0}  
\FAProp(5.,9.95)(2.,9.95)(0.,){Straight}{0}
\FAProp(5.,10.05)(2.,10.05)(0.,){Straight}{0}
\FAProp(5.,9.9)(2.,9.9)(0.,){Straight}{0}
\FAProp(5.,10.1)(2.,10.1)(0.,){Straight}{0}
\FAProp(5.,9.85)(2.,9.85)(0.,){Straight}{0}
\FAProp(5.,10.15)(2.,10.15)(0.,){Straight}{0}
\FAProp(5.,9.8)(2.,9.8)(0.,){Straight}{0}
\FAProp(5.,10.2)(2.,10.2)(0.,){Straight}{0}
\FAProp(7.5,5.)(6.,2.)(0.,){Straight}{0}  \FAProp(12.5,5.)(14.,2.)(0.,){Straight}{0}  

\end{feynartspicture}%

\par\end{centering}

\caption{a) wheel with four spokes ~ b) one-mass hexagon}

\end{figure}

\section{Conclusions}
In this article we have presented explicit algorithms for symbolic computation of iterated integrals on moduli spaces $\Mod_{0,n+3}$
of curves of genus $0$ with $n+3$ ordered marked points, based on \cite{Bro2}. These algorithms include the total differential of these functions, computation of primitives and 
the exact computation of limits at arguments equal to 0 and 1. The algorithms are formulated by use of operations on an explicit model for the reduced bar construction on $\Mod_{0,n+3}$ in terms of cubical 
coordinates $x_i$. In this formulation, the algorithms are well suited for an implementation on a computer. We have furthermore presented an algorithm
for the symbol map, out of  which the vector space of homotopy invariant iterated integrals on $\Mod_{0,n+3}$ can be constructed.

We expect the algorithms to apply to a variety of problems in theoretical physics and pure mathematics. Here we have concentrated on two main applications. 
As a first application, we have considered the computation of periods on $\Mod_{0,n+3}$, for which our algorithms are readily applicable. Secondly, we have discussed 
the computation of a class of Feynman integrals by the method of \cite{Bro4, Bog1}. In this approach, the Feynman integral is mapped to an integral on the moduli space, 
where our algorithms are applied to compute a single integration. We have presented an explicit procedure for the required change of variables from Schwinger parameters to 
cubical coordinates. A further procedure maps the result of the integration back to iterated integrals in terms of Schwinger parameters, and this process can be iterated. Using an implementation of our algorithms based on Maple, 
we have computed examples of such applications. As a third type of application, 
we have briefly discussed an approach for the expansion of generalized hypergeometric functions.

\end{document}